\documentclass[twocolumn]{aastex63}
\usepackage{graphicx}
\usepackage{upgreek}

\usepackage{xcolor}
\received{XXX}
\revised{YYY}
\accepted{ZZZ}

\newcommand{\chimepsr}{CHIME/Pulsar}
\newcommand{\presto}{PRESTO}
\newcommand{\psrchive}{PSRCHIVE}
\newcommand{\dspsr}{DSPSR}
\newcommand{\psrdada}{PSRDADA}
\newcommand{\tempo}{TEMPO}
\newcommand{\tempotwo}{TEMPO2}
\newcommand{\dmunits}{$\mathrm{pc\,cm^{-3}}$}

\newcommand{\ubc}{Department of Physics \& Astronomy, University of British Columbia, 6224 Agricultural Road, Vancouver, BC V6T 1Z1, Canada}
\newcommand{\mcgilldep}{Department of Physics, McGill University, 3600 rue University, Montr\'eal, QC H3A 2T8, Canada}
\newcommand{\mcgillsi}{McGill Space Institute, McGill University, 3550 rue University, Montr\'eal, QC H3A 2A7, Canada}
\newcommand{\mitkavli}{MIT Kavli Institute for Astrophysics and Space Research, Massachusetts Institute of Technology, 77 Massachusetts Ave, Cambridge, MA 02139, USA}
\newcommand{\mitdep}{Department of Physics, Massachusetts Institute of Technology, 77 Massachusetts Ave, Cambridge, MA 02139, USA}
\newcommand{\cita}{Canadian Institute for Theoretical Astrophysics, 60 St. George Street, Toronto, ON M5S 3H8, Canada}
\newcommand{\dunlap}{Dunlap Institute for Astronomy \& Astrophysics, University of Toronto, 50 St. George Street, Toronto, ON M5S 3H4, Canada}
\newcommand{\ubcok}{Department of Computer Science, Math, Physics, and Statistics, University of British Columbia, 3187 University Way, Kelowna, BC V1V 1V7, Canada}
\newcommand{\nrc}{National Research Council Canada, Herzberg Research Centre for Astronomy and Astrophysics, Domionion Radio Astrophysical Observatory, PO Box 248, Penticton BC V2A 6J9, Canada}

\begin{document}
\title{The CHIME Pulsar Project: System Overview}

\author{CHIME/Pulsar Collaboration}
\noaffiliation{}

\author[0000-0001-6523-9029]{M.~Amiri}
\affiliation{\ubc}

\author[0000-0003-3772-2798]{K.~M.~Bandura}
\affiliation{CSEE, West Virginia University, Morgantown, WV 26505, USA}
\affiliation{Center for Gravitational Waves and Cosmology, West Virginia University, Morgantown, WV 26505, USA}

\author[0000-0001-8537-9299]{P.~J.~Boyle}
\affiliation{\mcgilldep}
\affiliation{\mcgillsi}

\author[0000-0002-1800-8233]{C.~Brar}
\affiliation{\mcgilldep}
\affiliation{\mcgillsi}

\author[0000-0001-6509-8430]{J.-F.~Cliche}
\affiliation{\mcgilldep}
\affiliation{\mcgillsi}

\author[0000-0002-1529-5169]{K.~Crowter}
\affiliation{\ubc}

\author[0000-0003-2319-9676]{D.~Cubranic}
\affiliation{\ubc}

\author[0000-0002-6664-965X]{P.~B.~Demorest}
\affiliation{National Radio Astronomy Observatory, P.O. Box O, Socorro, NM 87801 USA}

\author[0000-0003-2381-9804]{N.~T.~Denman}
\affiliation{Central Development Laboratory, National Radio Astronomy Observatory, 1180 Boxwood Estate Road, Charlottesville, VA USA 22903}

\author[0000-0001-7166-6422]{M.~Dobbs}
\affiliation{\mcgilldep}
\affiliation{\mcgillsi}

\author[0000-0003-4098-5222]{F.~Q.~Dong}
\affiliation{\ubc}

\author[0000-0002-6899-1176]{M.~Fandino}
\affiliation{\ubc}

\author[0000-0001-8384-5049]{E.~Fonseca}
\affiliation{\mcgilldep}
\affiliation{\mcgillsi}

\author[0000-0003-1884-348X]{D.~C.~Good}
\affiliation{\ubc}

\author[0000-0002-1760-0868]{M.~Halpern}
\affiliation{\ubc}

\author[0000-0001-7301-5666]{A.~S.~Hill}
\affiliation{\ubcok}
\affiliation{\nrc}

\author{C.~H\"ofer}
\affiliation{\ubc}

\author[0000-0001-9345-0307]{V.~M.~Kaspi}
\affiliation{\mcgilldep}
\affiliation{\mcgillsi}

\author{T.~L.~Landecker}
\affiliation{\nrc}

\author[0000-0002-4209-7408]{C.~Leung}
\affiliation{\mitkavli}
\affiliation{\mitdep}

\author[0000-0001-7453-4273]{H.-H.~Lin}
\affiliation{\cita}
\affiliation{Max Planck Institute for Radio Astronomy, Auf dem Huegel 69, 53121 Bonn, Germany}

\author[0000-0001-5373-5914]{J.~Luo}
\affiliation{\cita}

\author[0000-0002-4279-6946]{K.~W.~Masui}
\affiliation{\mitkavli}
\affiliation{\mitdep}

\author[0000-0001-7453-4273]{J.~W.~McKee}
\affiliation{\cita}

\author[0000-0002-0772-9326]{J.~Mena-Parra}
\affiliation{\mitkavli}

\author[0000-0001-8845-1225]{B.~W.~Meyers}
\affiliation{\ubc}

\author[0000-0002-2551-7554]{D.~Michilli}
\affiliation{\mcgilldep}
\affiliation{\mcgillsi}

\author[0000-0002-9225-9428]{A.~Naidu}
\affiliation{\mcgilldep}
\affiliation{\mcgillsi}

\author[0000-0002-7333-5552]{L.~Newburgh}
\affiliation{Department of Physics, Yale University, New Haven, CT 06520, USA}

\author[0000-0002-3616-5160]{C.~Ng}
\affiliation{\dunlap}

\author{C.~Patel}
\affiliation{\dunlap}
\affiliation{\mcgilldep}

\author{T.~Pinsonneault-Marotte}
\affiliation{\ubc}

\author[0000-0001-5799-9714]{S.~M.~Ransom}
\affiliation{National Radio Astronomy Observatory, 520 Edgemont Rd., Charlottesville, VA 22903, USA}

\author[0000-0003-3463-7918]{A.~Renard}
\affiliation{\dunlap}

\author[0000-0002-7374-7119]{P.~Scholz}
\affiliation{\dunlap}

\author[0000-0002-4543-4588]{J.~R.~Shaw}
\affiliation{\ubc}

\author[0000-0002-1235-4485]{A.~E.~Sikora}
\affiliation{\mcgilldep}
\affiliation{\mcgillsi}

\author[0000-0001-9784-8670]{I.~H.~Stairs}
\affiliation{\ubc}

\author[0000-0001-7509-0117]{C.~M.~Tan}
\affiliation{\mcgilldep}
\affiliation{\mcgillsi}

\author[0000-0003-2548-2926]{S.~P.~Tendulkar}
\affiliation{\mcgilldep}
\affiliation{\mcgillsi}

\author[0000-0002-7436-2325]{I.~Tretyakov}
\affiliation{\dunlap}
\affiliation{Department of Physics, University of Toronto, Toronto, Ontario, M5S 3H4, Canada}

\author[0000-0003-4535-9378]{K.~Vanderlinde}
\affiliation{David A. Dunlap Department of Astronomy \& Astrophysics, University of Toronto, 50 St. George Street, Toronto, ON M5S 3H4, Canada}
\affiliation{\dunlap}

\author[0000-0002-1491-3738]{H.~Wang}
\affiliation{\mitkavli}
\affiliation{\mitdep}

\author[0000-0002-2472-6485]{X.~Wang}
\affiliation{School of Physics and Astronomy, Sun Yat-sen University, 2 Daxue Road, Zhuhai, China}

\correspondingauthor{A. Naidu}
\email{arun.naidu@mail.mcgill.ca}

\shorttitle{The CHIME Pulsar Project: System Overview}
\shortauthors{CHIME/Pulsar Collaboration et al.}

\begin{abstract}
We present the design, implementation, and performance of the digital pulsar observing system constructed for the Canadian Hydrogen Intensity Mapping Experiment (CHIME). Using accelerated computing, this system processes independent, digitally-steered beams formed by the CHIME correlator to simultaneously observe up to 10 radio pulsars and transient sources. Each of these independent streams are processed by the \chimepsr{} backend system which can coherently dedisperse, in real time, up to dispersion measure values of 2500\,\dmunits{}. The tracking beams and real-time analysis system are autonomously controlled by a priority-based algorithm that schedules both known sources and positions of interest for observation with observing cadences as rapid as one day. Given the distribution of known pulsars and radio-transient sources, and the dynamic scheduling, the \chimepsr{} system can monitor 400--500 positions once per sidereal day and observe most sources with declinations greater than $-20^\circ$ once every $\sim$4 weeks. We also discuss the extensive science program enabled through the current modes of data acquisition for \chimepsr{} that centers on timing and searching experiments. 

\end{abstract}


\section{Introduction}
The Canadian Hydrogen Intensity Mapping Experiment (CHIME\footnote{\url{https://chime-experiment.ca}}) is a radio interferometer operating in the 400--800 MHz frequency range, located at the Dominion Radio Astrophysical Observatory (DRAO) in Penticton, British Columbia, Canada. CHIME is a wide-field transit telescope with no physical-slew capabilities, with a primary design goal to study dark energy through the measurement of the evolution of baryon acoustic oscillations across the redshift range of $0.8 < z < 2.5$. This measurement will be accomplished by mapping neutral-hydrogen emission as a function of frequency across the entire Northern sky. From initial conception, it was realized that the CHIME telescope is also highly suitable for high-cadence observations of radio pulsars, while later planning and development was undertaken to enable untargeted searches for fast radio bursts \citep[FRBs; see e.g.][]{lbm+07,tsb+13,phl19}. To date, several telescope backends have been developed for CHIME that perform pulsar and FRB observations, as well as very-long baseline interferometry (VLBI) and forthcoming measurements of neutral-hydrogen absorption systems \citep{yzp14}. In this work, we describe the design, performance, and motivation of a pulsar-timing system constructed for CHIME, hereafter referred to as \chimepsr{}.

CHIME and the \chimepsr{} system differ from other radio observatories in two distinct ways: \chimepsr{} observes up to 10 different celestial positions at any instant in time, capable of acquiring spectro-temporal data in two different acquisition modes commonly used in pulsar and FRB astronomy; and \chimepsr{} is designed to observe continuously with the CHIME FX correlator, and commensally with other CHIME backends. These two capabilities allow \chimepsr{} to observe 400--500 known pulsars per day, and nearly all known sources in the the CHIME field of view (i.e., $\delta > -20^\circ$) within one month. CHIME therefore acts as an automated, high-cadence observatory for pulsar science.

Until recently, only a small fraction of the $\sim$2800 known radio pulsars was observed regularly using other radio facilities due to competitive and limited telescope resources. Nonetheless, high-cadence observations of many millisecond pulsars (MSPs) have been projected to: boost sensitivity towards nearby, resolvable sources of nanohertz-frequency gravitational waves (GWs) emitted from merging galaxies \citep[e.g.][]{cal+14}; resolve abrupt temporal variations in dispersion properties that reflect a complex interstellar medium \citep[ISM; e.g., ][]{leg+18}; and constrain the equation of state of ultra-dense matter \citep[e.g.][]{dpr+10}. Furthermore, high precision timing of an array of many stable MSPs is being used to search for, and directly detect, the low frequency GW background \citep[e.g.][]{vlh+16,pdd+19}, a concept commonly referred to as a ``pulsar timing array" (PTA).

Renewed and frequent tracking of historically low-priority sources will likely resolve dynamical processes affecting pulsar rotation, such as unidentified long-period orbital motion \citep{kkn+16,bjs+16,ant20} and many pulsar ``glitches" \citep[e.g.][]{elsk11}. The \chimepsr{} system is also effective at monitoring sporadically emitting pulsars and other transient events that can provide great insight into pulsar emission intermittency and astrophysical coherent radio emission mechanisms more broadly. The ability of \chimepsr{} to monitor the entire Northern-hemisphere pulsar population on a regular basis is therefore expected to yield unprecedented access to a large number of timing-based phenomena that previously could not be resolved elsewhere.

A brief introduction to the \chimepsr{} project has been presented by \citet{Ng2018}. Here we provide a more detailed overview of the \chimepsr{} system, its placement within the CHIME infrastructure, the performance of \chimepsr{}, and science topics to be addressed with the \chimepsr{} system. In Section \ref{sec:hardware}, we describe the hardware components of the CHIME instruments and the pulsar-timing backend. In Section \ref{sec:software}, we provide an overview of the software suite developed to make pulsar-timing measurements possible with CHIME. In Section \ref{sec:performance}, we outline the current performance of the \chimepsr{} system and provide a sample of preliminary results in demonstration of CHIME's capabilities as a pulsar-timing observatory. In Section \ref{sec:science}, we discuss various scientific opportunities enabled by the \chimepsr{} system. Finally, we summarize this work and its anticipated science outcomes in Section \ref{sec:conclusions}.

\section{Hardware}
\label{sec:hardware}
\begin{figure}[!htbp]
\includegraphics[width=\columnwidth]{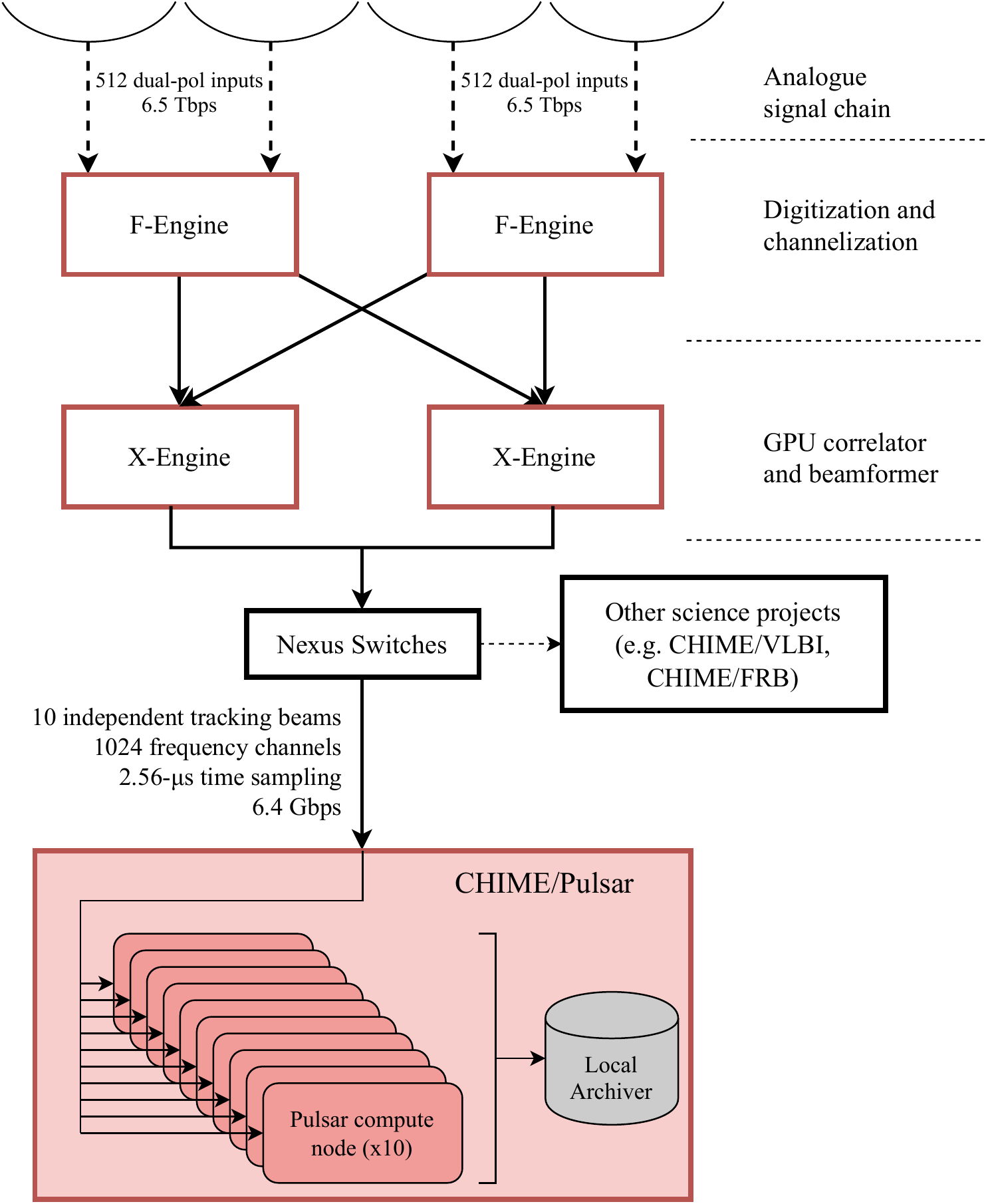}
\caption{Schematic of the CHIME telescope signal path. The telescope structure (four cylinders, here black arcs), the correlator (F- and X-Engines), and the \chimepsr{} backend are shown. The dashed black lines represent coaxial cables carrying the analogue signal from the 256 feeds on each cylinder to the F-Engines in the East/West receiver enclosures, situated below the telescope. The total input data rate into the F-Engine is 13\,Tbps. The solid black lines depict digitized data carried through copper and/or fiber-optic cables. Networking devices between the F- and X-Engines are not shown. The X-Engine (GPU-based correlation and beamforming) is housed in two shipping containers adjacent to the cylinders. The \chimepsr{} backend (shaded red) is housed in a shielded room in the DRAO building.  The data rate into the \chimepsr{} backend is 6.4\,Gbps.}
\label{fig:system_schematic}
\end{figure}

CHIME consists of 1024 dual-polarization radio-frequency inputs coupled to a correlator and then several independent backend digital instruments. Here we briefly describe the major signal-chain components relevant to the \chimepsr{} project. See Figure~\ref{fig:system_schematic} for a schematic diagram of the signal chain from the receivers to the \chimepsr{} backend.

\subsection{Telescope Structure and Receiver}
The CHIME telescope consists of four 5-m focal length parabolic cylinders that are 20\,m wide and 100\,m long. 
The cylinders are oriented North-South, each mounted with 256 dual-polarization 'cloverleaf' antennas \citep{Deng17} placed 30\,cm apart below a 68\,cm-wide groundplane at the parabolic focus.
Two low noise amplifiers are coupled directly to each antenna. 
These are coupled via 60\,m of low-loss coaxial cables to custom band-defining filter-amplifiers located in a radio-frequency shielded room containing analog-to-digital converters and the frequency channelization portion of the CHIME correlator.
Basic telescope properties are given in Table~\ref{tab:spec-chime}.

\begin{table}[htbp]
\centering
\caption{Telescope specifications.\label{tab:spec-chime}}
\begin{tabular}{ll}\hline
Field of view & 120$^{\circ}$ (N-S) \\
              & 1.3-2.5$^{\circ}$ (E-W) \\
Receiver noise temperature & $\sim$50\,K (nominal) \\
Frequency range & 400--800\,MHz \\
Polarization basis & orthogonal linear \\
Telescope longitude & $-119\degr 37' 25.238''$ \\
Telescope latitude & $49\degr 19' 14.553''$ \\
Telescope elevation\tablenotemark{a} & 547.918\,m \\
\hline 
\end{tabular}
\tablenotetext{a}{Above the GRS 80 ellipsoid.} 
\end{table}

\subsection{FX Correlator}
CHIME employs a hybrid FX correlator that consists of custom-built electronics and adapted commodity processing units.
The first stage, known as the F-Engine, is responsible for signal digitization and frequency separation, while the second stage, X-Engine, is responsible for spatial correlation and beamforming.

\subsubsection{F-Engine: Digitization and Channelization}
The amplified input signals are digitized and separated into 1024 frequency channels, resulting in a spectral resolution of 390\,kHz, by the F-Engine system. The details of the F-Engine are presented in \citet{ice1} and are discussed further in \citet{chimefrb}. In summary, the F-Engine consists of 128 ``ICE" FPGA-based motherboards where the signals from the 1024 dual-polarization receivers are digitized at 800 million samples per second with 8-bit precision. The baseband streams are channelized by a 4-tap polyphase filterbank into 1024 frequency channels. The positive frequency spectra of these channelized data are rounded to (4+4)-bit complex numbers after applying a programmable gain and phase offset, which halves the input data rate of 13.1\,Tbps to 6.5\,Tbps. The complex-voltage data are further re-organized to form 1024 data streams, which are transmitted via fibre-optic connection to the next processing stage.

\subsubsection{X-Engine: GPU Correlator and Beamformer}
The CHIME correlator X-Engine is a computing cluster consisting of 256 nodes hosting a total of 512 dual-chip AMD FirePro 9300 X2 graphics processing units (GPUs). These nodes are divided between two emission-shielded shipping containers located on the east side of the CHIME reflectors. Each X-Engine node processes four frequency channels -- one frequency per GPU chip -- for each of the 1024 dual-polarization receiver inputs. A set of GPU kernels process the input data to manipulate the CHIME baseband into a variety of data products required by downstream backends. The kernel used by \chimepsr{} system is discussed in Section ~\ref{sec:bf-cal}. The X-Engine data products are exported from the nodes over Gigabit Ethernet (GbE) to a set of commercially-available network switches, and then to various backend systems. A full description of the CHIME X-Engine and its capabilities is given by \cite{drv+20}.

\subsection{\chimepsr{} Backend}
The \chimepsr{} backend consists of ten independent compute nodes that are connected directly to the CHIME correlator through a series of network switches.
Each compute node consists of a Supermicro motherboard, an Intel Xeon E5-1650 central processing unit (CPU), 128\,GB of RAM, and a solid-state drive (SSD). Data are received through a 10-Gbps Intel Network Interface Controller (NIC) and is temporarily buffered in RAM before being processed through an FFT-based coherent dedispersion algorithm on a single, liquid-cooled NVIDIA Titan X GPU.
The resulting data are temporarily stored and further processed on a local, 60-TB data archiver. 

\subsection{Local Data-archiving Server}
The \chimepsr{} backend ultimately writes a variety of data products with variable file sizes several hundred times every sidereal day, at a time-averaged rate of 67\,Mbps.
In order to handle the output data rate, we built a local data-archive server using components similar to the hardware described above.
The archiving server employs two redundant arrays of independent disks (RAIDs) for secure, short-term storage of \chimepsr{} data products.
This short-term data archive can accommodate several months of continuous typical data acquisition, and is sufficient in the event that higher data-rate modes (see Section~\ref{sec:data_processsing}) become emphasized for a moderate time span.
These data are eventually transferred offsite via ground shipment of physical hard drives to \chimepsr{} institutions for uploading to multi-purpose computing facilities for offline processing and long-term preservation.
The primary long-term data archiving occurs on the Cedar cluster of Compute Canada\footnote{\url{https://www.computecanada.ca}}.

\section{Software}
\label{sec:software}
Here we describe the suite of software tools we have developed and implemented for rendering sky signals from the FX correlator into useful \chimepsr{} data products. Salient details of the \chimepsr{} backend output are given in Table~\ref{tab:spec-psr}.

\begin{table*}
\centering
\caption{\chimepsr{} data specifications.\label{tab:spec-psr}}
\begin{tabular}{ll}\hline
 Number of tied-array beams & 10 \\
 Beam width (FWHM) & $\sim$0.5$^{\circ}$ (at 400\,MHz) \\
                   & $\sim$0.25$^{\circ}$ (at 800\,MHz) \\
 Number of spectral channels & 1024 \\
 Frequency resolution & 390.625\,kHz \\
 Time resolution & 2.56\,$\upmu$s (baseband)\\
                 & 327.68\,$\upmu$s (output filterbank)\\
 Output data bit depth & 8 \\
 Number of profile bins\tablenotemark{a} & 256/512/1024/2048 (fold-mode)\\
 Output polarization state & Full Stokes (fold-mode) \\
                           & Total intensity (filterbank) \\
                           & X, Y (baseband) \\
 Pulsar data output rate\tablenotemark{b} & 67\,Mbps \\
\hline
\end{tabular}
\tablenotetext{a}{Configurable options set when scheduling observations.}
\tablenotetext{b}{Time-averaged over a typical observing day.}
\end{table*}

\subsection{Beamforming and Calibration}\label{sec:bf-cal}
We digitally form 10 dual-polarization tied-array beams within the X-Engine by summing all 1024 inputs phased to specified celestial coordinates. The 10 independent beams allow for tracking of 10 different sky positions simultaneously within the primary beam of CHIME. The re-pointing of each beam is performed over the network using the Representational State Transfer (REST) application programming interface, and typically occurs within milliseconds of command execution.

The complex gain calibration (amplitude and phase) of the input data is achieved via point-source calibration with bright continuum sources such as Cas~A, Cyg~A, and Tau~A (CHIME Collaboration et al., in prep.). During a bright continuum source transit, the visibility matrix for CHIME is treated as approximately the product of the complex gain and sky signal (as viewed through the beam response function). In this limit, computing an eigendecomposition of the visibility matrix allows us to determine the complex gain for each antenna at each frequency channel in the direction of the calibrator source. The eigenvectors corresponding to the two largest eigenvalues represent the complex gains for the X and Y polarizations of each input. After determining the eigenvectors, we also remove known interferometric phase based on the physical separation of the antennas by a process known as fringe-stopping \citep[e.g.][]{tms01}.

As part of the calibration process, data from poorly behaving antennas and contaminated frequencies are masked. A static list of inputs which are known to be malfunctioning are removed and this list is updated on a daily basis. Calibration solutions with more than 5\% of inputs deemed bad are flagged as poor quality, and the most recent, preceding solutions deemed adequate are instead used until the next calibrator transit occurs. While radio frequency interference (RFI) excision in CHIME data can be a serious challenge, some channels are known to always be contaminated, and these channels are masked in the calibration solutions.

During the ``pre-commissioning" period of CHIME operations (before August 2018), phase-only calibration was computed every few days. Since September 2018, daily amplitude and phase calibration solutions have been calculated. 

\subsection{Data packet format, transfer and assembly}\label{packet_format}
For each X-Engine GPU node, the resulting beamformed data from each of the four processed frequency channels are scaled down, encoded into (4+4)-bit complex numbers, and packetized in VLBI Data Interchange Format (VDIF)\footnote{\url{https://vlbi.org/vlbi-standards/vdif}}.
Each individual packet consists of 625 samples of one of the two polarizations for each of the four frequency channels and a VDIF header of 32 bytes, resulting in a total packet size of 5032\,bytes.
The corresponding pulsar data rate on each FX-correlator node for all 10 beams is 0.2516\,Gbps, with a total network load of $\sim$64\,Gbps for all pulsar beams.
Each individual pulsar-backend node receives $\sim$6.4\,Gbps of data from the correlator in the form of User Datagram Protocol (UDP) packets.   
The incoming data rate to each pulsar nodes corresponds to $1.6\times 10^5$ packets per second. We have optimized our network-capture algorithms in order to properly receive packets with negligible packet loss\footnote{The details about the optimizations employed are beyond the scope of this paper, but available upon request.}. The received packets are assembled and written to memory in preparation of further processing.

\begin{figure*}[!htbp]
\centering
\includegraphics[width=0.85\textwidth]{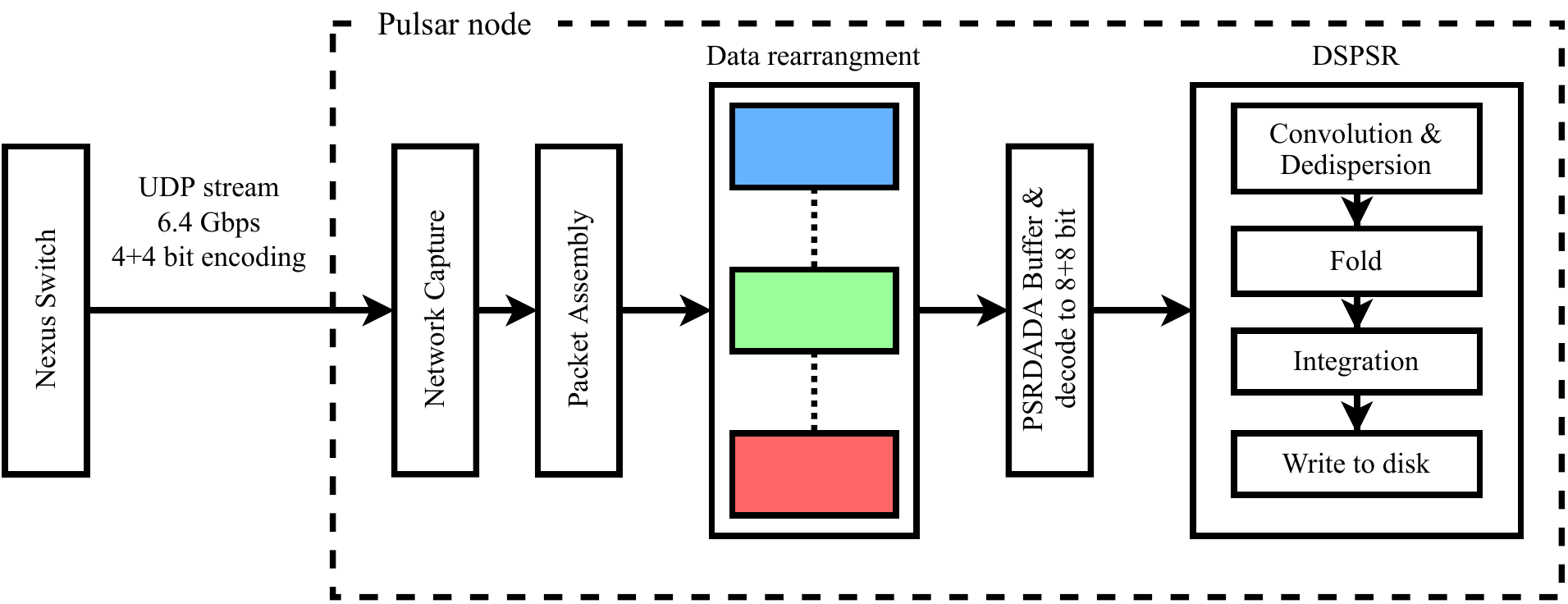}
\caption{A schematic illustrating the various operations on the data streams from the correlator in fold mode, with colors denoting frequency channels. The unordered UDP packets from the correlator are routed to the specified pulsar node where the data are organized in the assembler stage which involves both network capture and packet assembly. The assembled data, reordered into a polarization-frequency-time series in order of increasing index variation, are then passed on to the \dspsr{} process using \psrdada{} buffers.}
\label{fig:dspsr}
\end{figure*}

\subsection{Data processing}\label{sec:data_processsing}
The main purpose of the \chimepsr{} system is to autonomously monitor radio pulsars with regular cadence. The beamformed dual-polarization timeseries are coherently dedispersed and processed to form the data product defined by the scheduled observation configuration. Most observations occur in ``fold mode,'' where the dedispersed baseband data are manipulated to yield time-averaged pulse profiles for all 1024 frequency channels over a specified integration time, using a set of pulsar rotation parameters for coherent averaging of the signal. In most cases, the number of profile phase bins for a given fold-mode observation is 256, although this number is nominally defined by the science case and, therefore, dictated by the scheduling process. By default, 10-second integrations are formed for all fold-mode observations, although this could in principle be adjusted on a per-pulsar basis. When desired, observations may also be conducted in ``filterbank mode,'' where the dedispersed baseband timeseries are converted into a Stokes-$I$ data stream that is then downsampled in time. In special circumstances, it is scientifically worthwhile to directly record beamformed baseband data for specific pulsars and FRBs of interest. We have developed the appropriate software capabilities to enable all three types of observation with the \chimepsr{} system. The basic output data specifications of the \chimepsr{} system are given in Table~\ref{tab:spec-psr}.

\subsubsection{Fold mode}
For fold-mode observations, the \chimepsr{} backend uses the ``Digital Signal Processing for Pulsars'' (\dspsr{}) suite\footnote{\url{http://dspsr.sourceforge.net}}, an open source GPU-based library \citep{DSPSR}. Written in C++ and CUDA\footnote{\url{https://developer.nvidia.com/cuda-zone}}, \dspsr{} is a high performance, general-purpose tool for high-time-resolution radio pulsar studies using accelerated computing. \dspsr{} is used in many astrophysical applications \citep[e.g.][]{ksl12,msf+15,psb+16} and is extensively used as a part of real-time processing instrumentation for various telescopes around the world.

\dspsr{} has the ability to read from and write to data buffers created using the \psrdada{} library\footnote{\url{http://psrdada.sourceforge.net}}. \psrdada{} is an open source software that allows the creation of flexible and well managed ring buffers, with a variety of applications for piping data from process to ring buffer, and vice versa. In addition to enabling transfer of data between processes, \psrdada{} provides various utilities to manage data within the ring buffer itself.

For fold-mode observations, the incoming data streams from the CHIME X-Engine are sorted and arranged as a polarization-frequency-time series prior to entering the real-time processing stage. The assembled data are then passed onto the \dspsr{} process using \psrdada{} buffers. Even though it has the ability to accept data with (4+4)-bit encoding, \dspsr{} runs significantly faster when instead receiving data with (8+8)-bit encoding. We have thus modified the \psrdada{} software to convert data to (8+8)-bit encoding before passing them to \dspsr{} in order to circumvent this issue. Once the data are passed into the \dspsr{} GPU buffers, \dspsr{} executes several kernels that perform coherent dedispersion and timing-model folding of the data before writing to disk. A schematic of the processing chain for fold-mode observations is presented in Figure \ref{fig:dspsr}.

\begin{figure}[!htbp]
\centering
\includegraphics[width=\columnwidth]{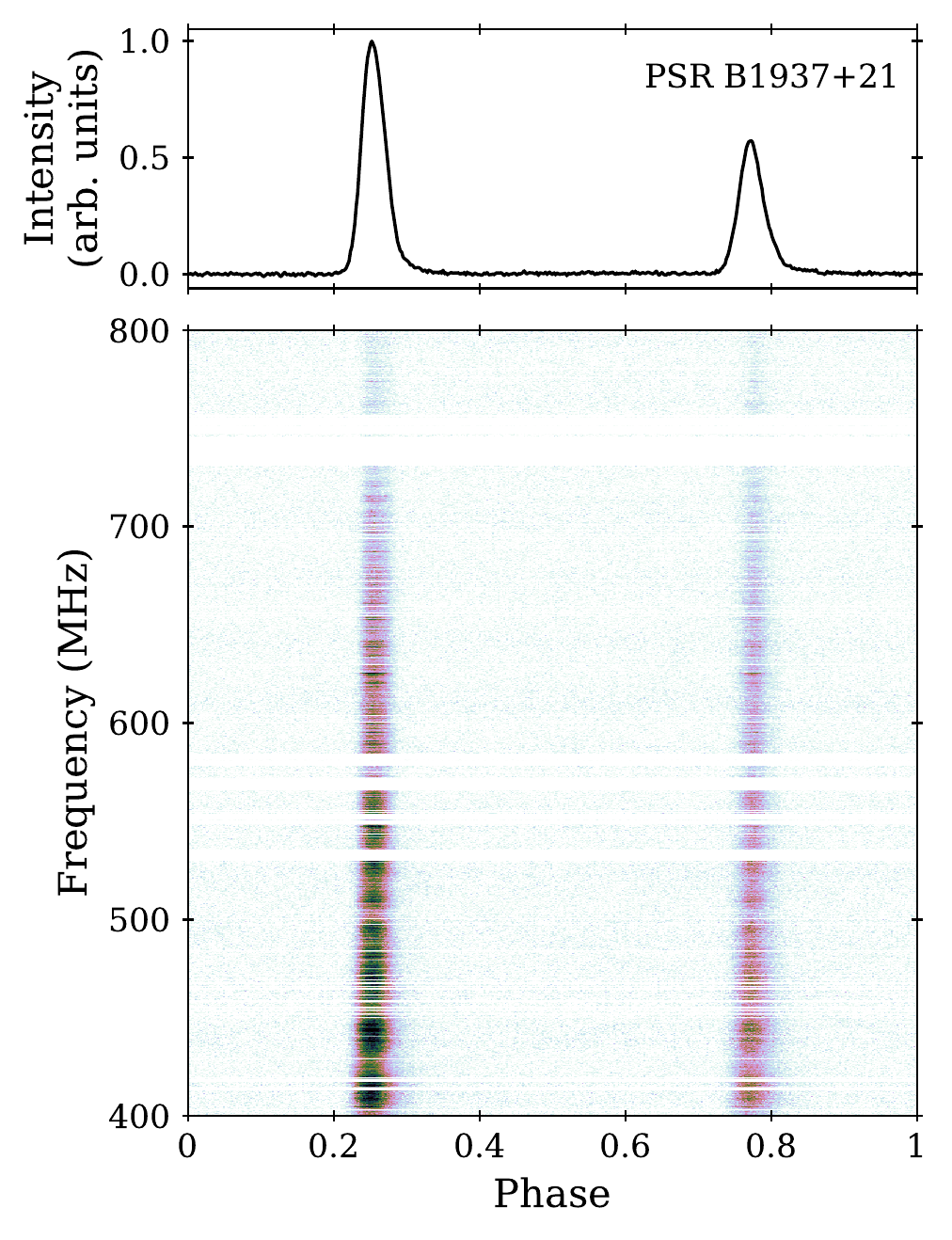}
\caption{A fold-mode observation of PSR~B1937+21. The bottom panel is the phase versus frequency (waterfall) plot, and the top panel is the mean intensity after summing all frequency channels (i.e. the integrated pulse profile). There is clear evidence of pulse broadening due to interstellar scattering visible in the folded profile and as a function of frequency across the 400\,MHz bandwidth.\label{fig:B1937}}
\end{figure}

Output fold-mode data products are written to structured files (``archives'') that are readable by the \psrchive{} software suite\footnote{\url{http://psrchive.sourceforge.net}} \citep{PSRCHIVE_hvm04,PSRCHIVE_vdo12} for offline processing and analysis. 
An example of the \chimepsr{} system fold-mode data from an observation of PSR~B1937+21 is given in Figure~\ref{fig:B1937}.

\subsubsection{Filterbank mode}
\label{subsec:filterbank}
The CHIME/FRB system is discovering many repeating FRBs \citep{chimefrb_second_repeater,chimefrb_repeaters19,fab+20}, as well as single pulses from unknown radio pulsars and ``rotating radio transients" \citep[RRATs;][]{mll+06}. 
Some fraction of the \chimepsr{} observing time is thus expected to be used for follow-up observations of these sources. 
To observe repeating FRB sources, the pulsar backend must be able to record coherently dedispersed high-time-resolution filterbank data. 
Since these extragalactic sources are known to have DMs upwards of 1000\,\dmunits{} it is also essential that the backend can support such DM ranges. 

\begin{figure*}
\centering
\includegraphics[width=0.85\textwidth]{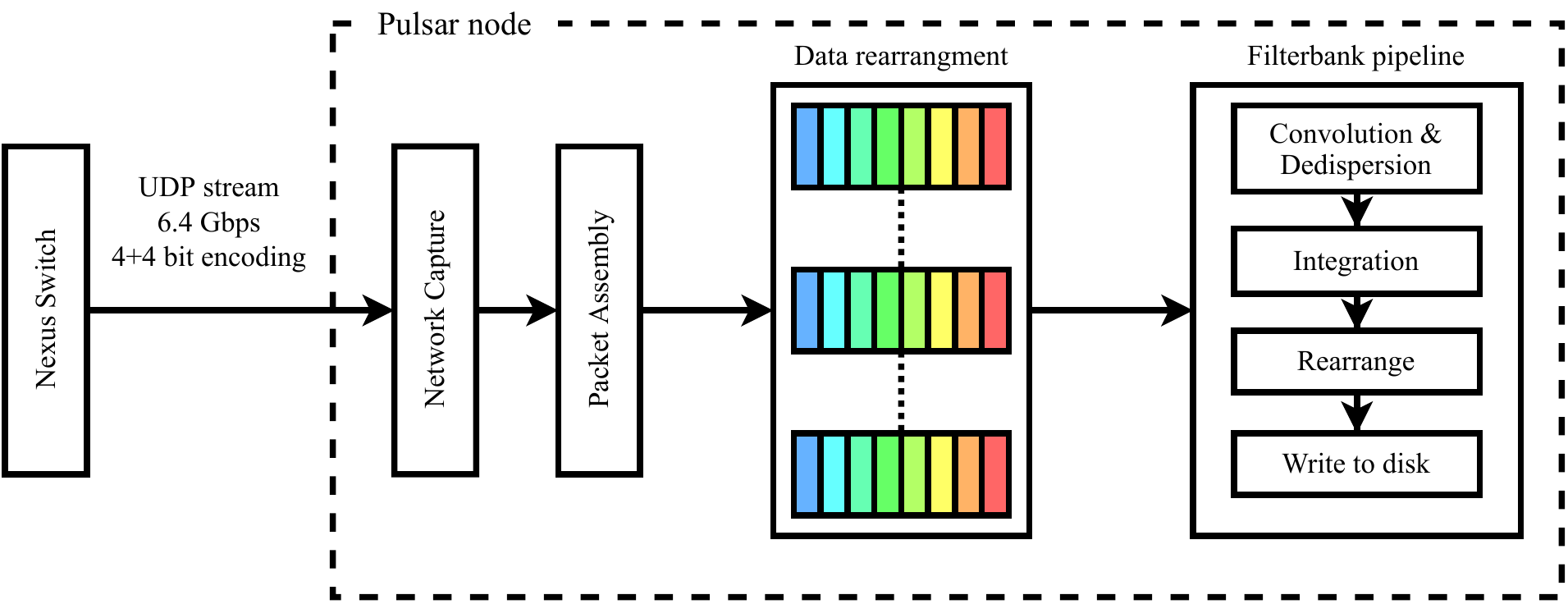}
\caption{A schematic illustrating the various operations of the data streams from the correlator in the filterbank mode, with colors denoting frequency channels. As in the fold-mode diagram, the scrambled UDP packets from the correlator are routed to the specified pulsar node where the data are sorted and organized prior to real-time processing. However, the data stream is instead assembled into a polarization-time-frequency series and in order of increasing index variation, for reasons discussed in Section \ref{subsec:filterbank}. The assembled data are then passed on to the coherent dedispersion pipeline process and then to filterbank generation and output.}
\label{fig:filterbank}
\end{figure*}

The readily available \dspsr{} filterbank module \textit{digifil} cannot support real-time performance at such large DMs at the CHIME operating frequency range. 
To overcome this, we have developed a new codebase for the filterbank mode with the ability to dedisperse incoming data to a maximum DM of 3300\,\dmunits{} in real-time based on work by \cite{njm+}.
The significant change in our implementation is the input data format. Based on benchmark tests on FFT performance, the efficiency of the GPUs can be improved by a factor of five if the input data format is changed from polarization-frequency-time series to polarization-time-frequency series as shown in Figure~\ref{fig:filterbank}.

The received UDP packets are arranged into a polarization-time-frequency series and are passed on to the GPU for coherent dedispersion at a user-specified DM. 
This dedispersed data are then passed on to the CPU for downsampling, and then manipulated into a community-established filterbank format before writing to disk. 
All the code for the CPU processes after GPU kernel execution are written in Intel AVX2 instrinsics.

\subsection{Scheduler and web monitors}
\label{subsec:scheduler}

An automated scheduler is necessary to make maximal use of the continuous operation of CHIME for pulsar observations. The transit nature of the telescope means that sources are set to be observed only when they are within $1.75^\circ$ of the meridian. This results in certain parts of the sky having more sources than the number of available beams. To address this, we have developed a probabilistic scheduling algorithm that employs user-defined weights to prioritize certain sources, or positions of interest, over others. This priority system is used to resolve scheduling conflicts. There are five base weight values, where the highest weight is assigned to MSPs currently used in a PTA for gravitational-wave detection, as well as other interesting pulsars (e.g., relativistic pulsar-binary systems) such that they will be observed every day. The remaining four weighting values are defined such that each additional tier has twice the weight of the previous tier. There is also a feedback mechanism that reduces the priority associated with a source (as long as it is not assigned the highest priority level) based on the number of days it has been observed over the previous 14 days. The weight of a previously scheduled source will be reduced by a factor equivalent to the number of days observed over the period.

The scheduler adds the next source to the schedule through an iterative process. It looks for sources whose transit time began within 7 minutes of the end of the transit time of the previous source for each one of the ten beams. In cases where two or more sources fall within the time-frame, the scheduler will automatically select the source with a weight that corresponds to being observed daily and starting time of transit being closest to the end of the transit time of the previous source. If none of the sources have such weighting, each source will be given a probability equivalent to its weighting divided by the weighting of all sources in conflict and a source will be randomly selected with the determined probabilities.

\begin{figure}
    \centering
    \includegraphics[width=\linewidth]{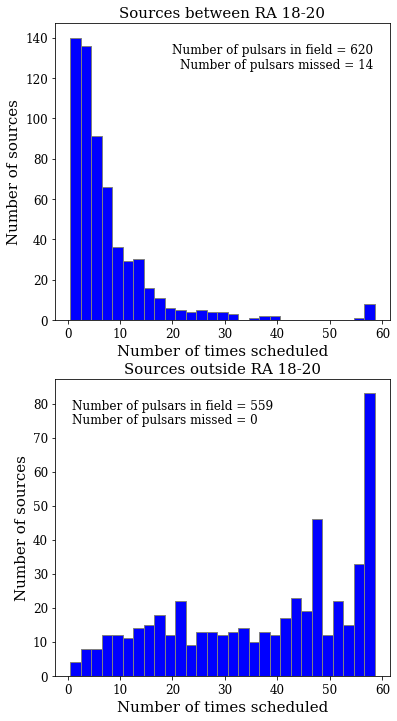}
    \caption{Histograms showing the rate that each source is observed over a period of 60 days, for sources located at Right Ascensions (RAs) of between 18-20 and for those outside of the RA range. The histograms is binned in units of two, starting with the first bar representing number of sources scheduled for either 1 or 2 times over the past 2 months. The number of sources available for each RA range and the number of sources not scheduled over the 60-day period are indicated on the plots.}
    \label{fig:scheduler_hist}
\end{figure}

In order to evaluate the effectiveness of the scheduler, we examined the number of times each source is scheduled over a 60-day period. Figure~\ref{fig:scheduler_hist} shows the histograms of sources and how often they are scheduled. We found that when the telescope is not transiting the Galactic plane, the scheduler is highly effective in cycling through all sources, with more than half of the sources being observed on near-daily cadence (i.e., scheduled $>30$ times over 60 days), with only about 10 of the 559 sources having a cadence of 2 weeks or more (i.e., scheduled $<4$ times over 60 days). On the more crowded field of RAs between 18-20 hours, sources on average will have a longer cadence, with less than 20 sources scheduled with near-daily cadence and about 290 out of the 620 sources having an observing cadence of 2 weeks or longer, among them 14 sources that were not scheduled at all. While this circumstance is less than ideal, it nonetheless demonstrates that the scheduler is able to cycle through most of the sources even with a large number of conflicts.

We also allow for flexibility and autonomous schedule interruption in the case of special target of opportunity events. 
For example, if there is a candidate alert from the CHIME/FRB system or if a radio magnetar is reported to burst, we can override the standard schedule to point at the location as soon as it is visible. 
It is also possible to grid around a given source position simultaneously by employing multiple tracking beams. 

Multiple web-based tools are in place to actively monitor the continuous operation, including a live data-product viewer and a Grafana\footnote{https://grafana.com} visualization platform for displaying system, transmission and data-storage metrics collected using the Prometheus toolkit\footnote{https://prometheus.io}.

\section{Backend Performance} \label{sec:performance}

The software infrastructure and scheduling algorithm developed for the \chimepsr{} system, described in Section \ref{sec:software}, allow for priority-based observations of many known pulsars in the Northern hemisphere on a continuous basis. On average, the \chimepsr{} system observes roughly 400--500 pulsars per day, and monitors all known sources with declinations greater than $-20^\circ$ at least once every $\sim$4 weeks. Per-source observing cadence varies as a function of source right ascension due to the heterogeneous nature of the pulsar sky distribution, as discussed in Section \ref{subsec:scheduler}, though the \chimepsr{} system nonetheless observes nearly all pulsars at least once within a several-week timescale. We demonstrate the observing capabilities of the \chimepsr{} system by describing backend performance during the initial 1.5 years of CHIME operation (July 2018 -- November 2019).

\subsection{System up-time}
During the ``pre-commissioning" period of CHIME operations, from July 2018 to February 2019, we gradually increased the number of pulsar-backend nodes to the current ten available and employed a monthly operation cycle with a week of software deployment for new features between cycles. The \chimepsr{} system typically had an up-time of $\sim$70\% during the pre-commissioning phase of operation (including downtime due to software-deployment weeks). As expected, the on-sky observing time has increased to $\sim$100\% as the commissioning period continues and transitions to the ``full-science" mode of CHIME operation. 
As of publication, the \chimepsr{} backend is continuously acquiring data alongside the various backends currently deployed for the CHIME telescope.

\begin{figure*}[htbp]
\centering
\includegraphics[width=\linewidth]{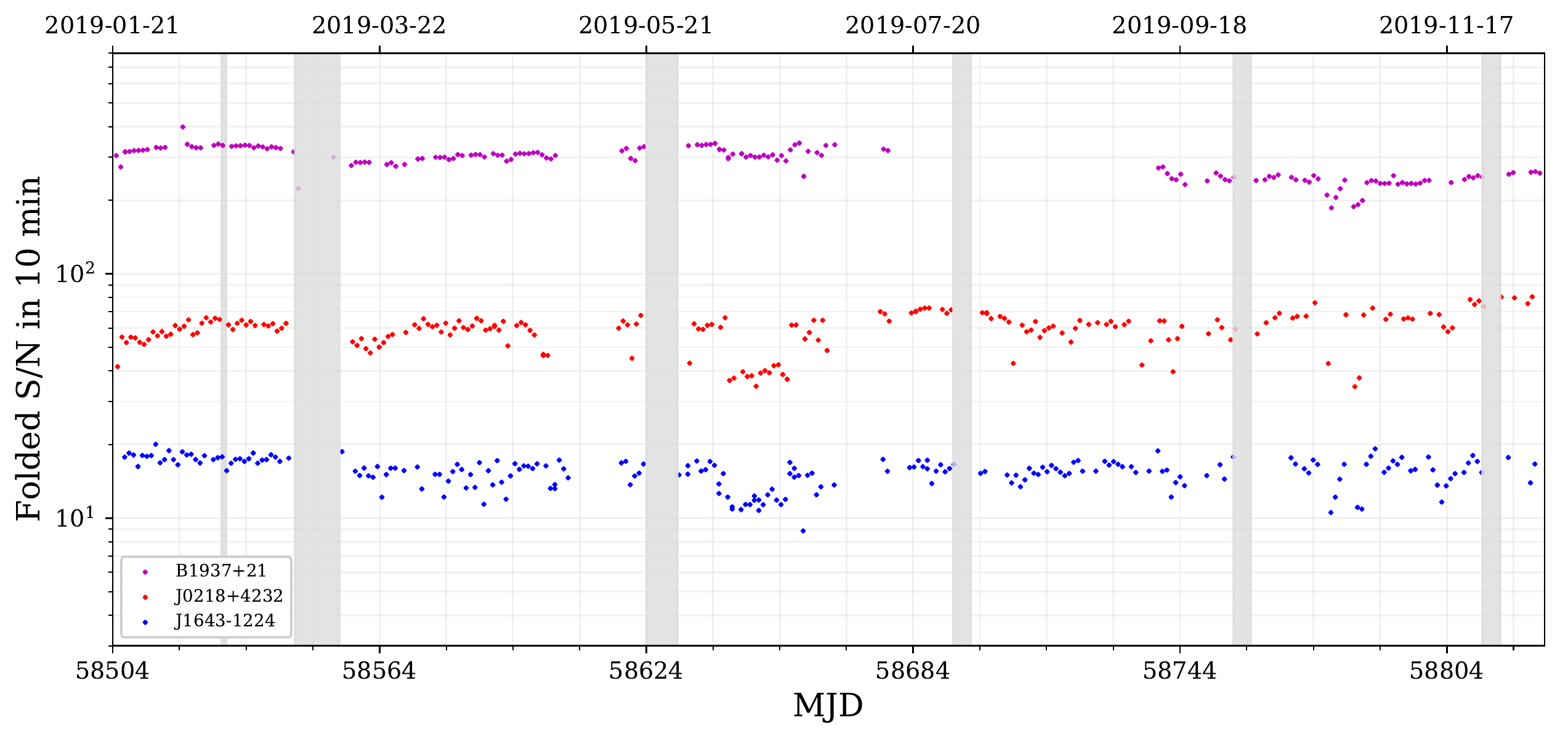}
\caption{Observed S/N for three pulsars (PSRs J1643$-$1224, J0218+4232 and B1937+21) in 2019. Vertical grey bands correspond to correlator-software deployment weeks where no science-quality data were recorded.}
\label{fig:dailyS/N}
\end{figure*}

Figure~\ref{fig:dailyS/N} shows a selection of MSPs and their daily signal-to-noise ratio (S/N) variations as observed by \chimepsr{} backend over the year of 2019 commissioning period. 
These S/N variations indicate that the system is stable during the commissioning period. 
Some degree of S/N variation remains, though we believe this is due to imperfect excision of the time-varying RFI environment, imperfections in the calibration solutions, and potentially scintillation effects towards low-DM pulsars. 

\subsection{Radio frequency interference mitigation}
Frequency channels affected by bright RFI (notably the LTE cellular network band between $\sim$730-755\,MHz and several digital television bands between 500-600\,MHz) are identified from the amplitude and phase calibration solutions provided for each antenna and frequency channel (see Section~\ref{sec:bf-cal}).
Since these solutions are calculated and applied on a daily basis, the particular frequency channels that are masked can vary on similar timescales, but the RFI environment is generally stable.

\begin{figure*}
\centering
\includegraphics[width=0.47\textwidth]{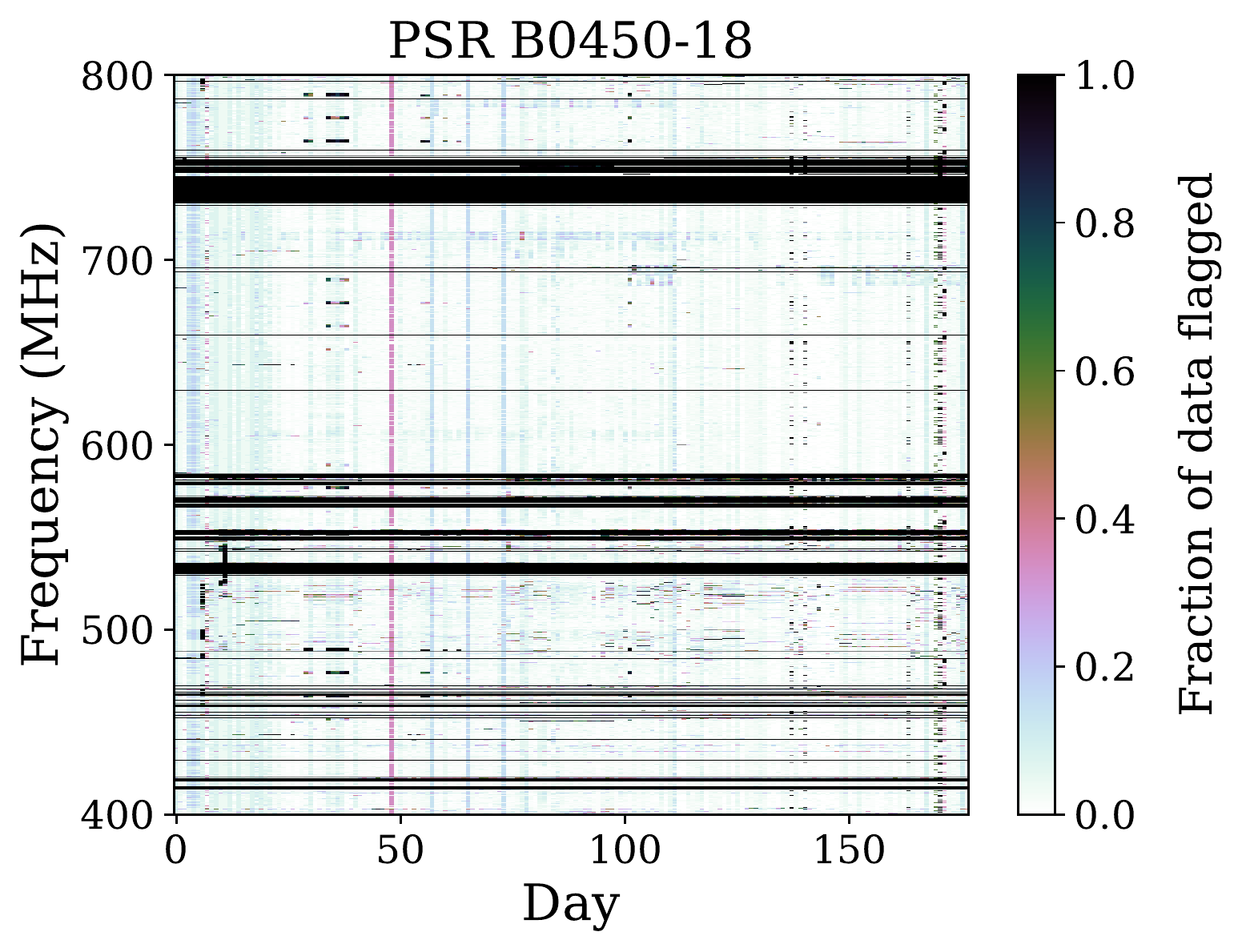}
\includegraphics[width=0.47\textwidth]{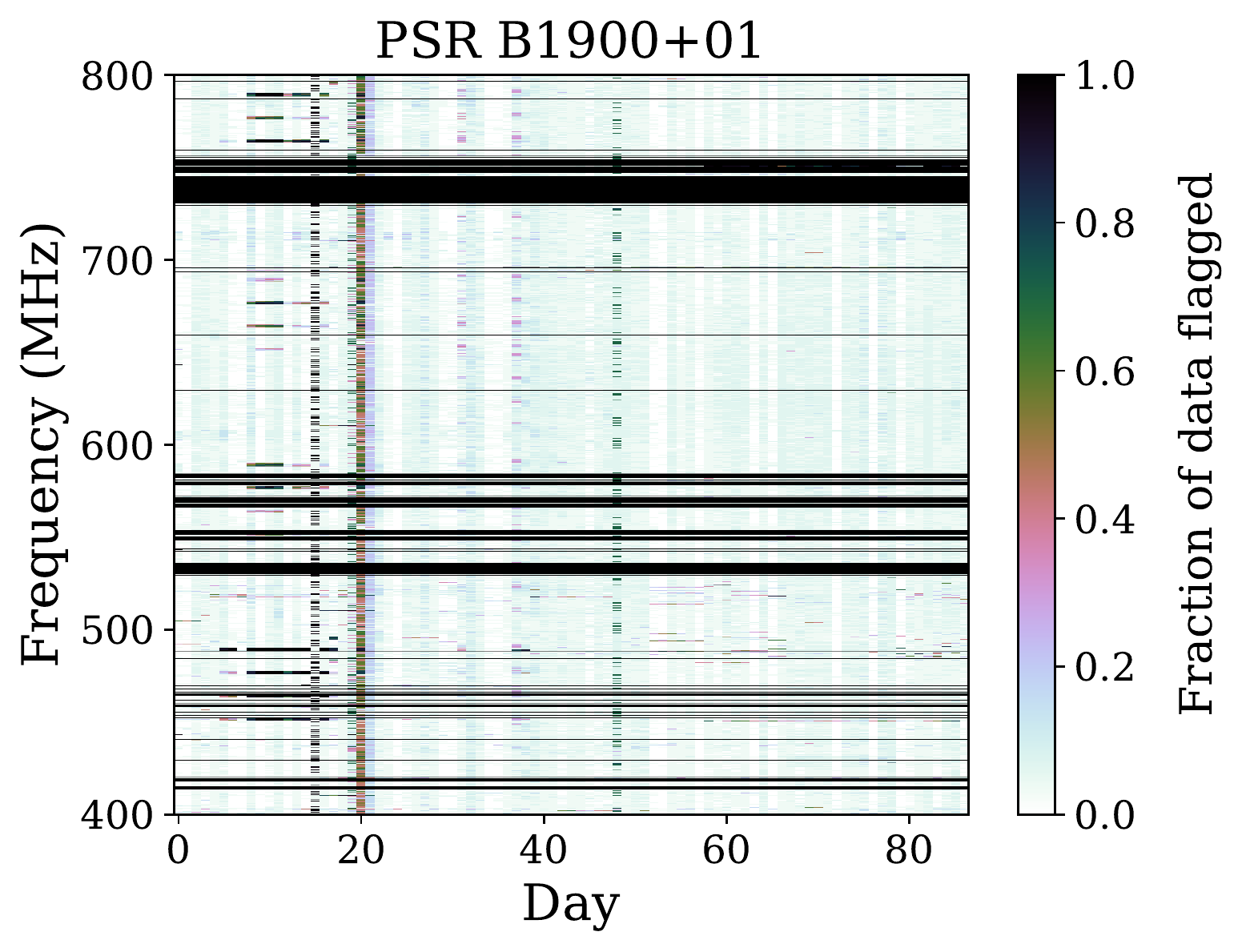}
\includegraphics[width=0.47\textwidth]{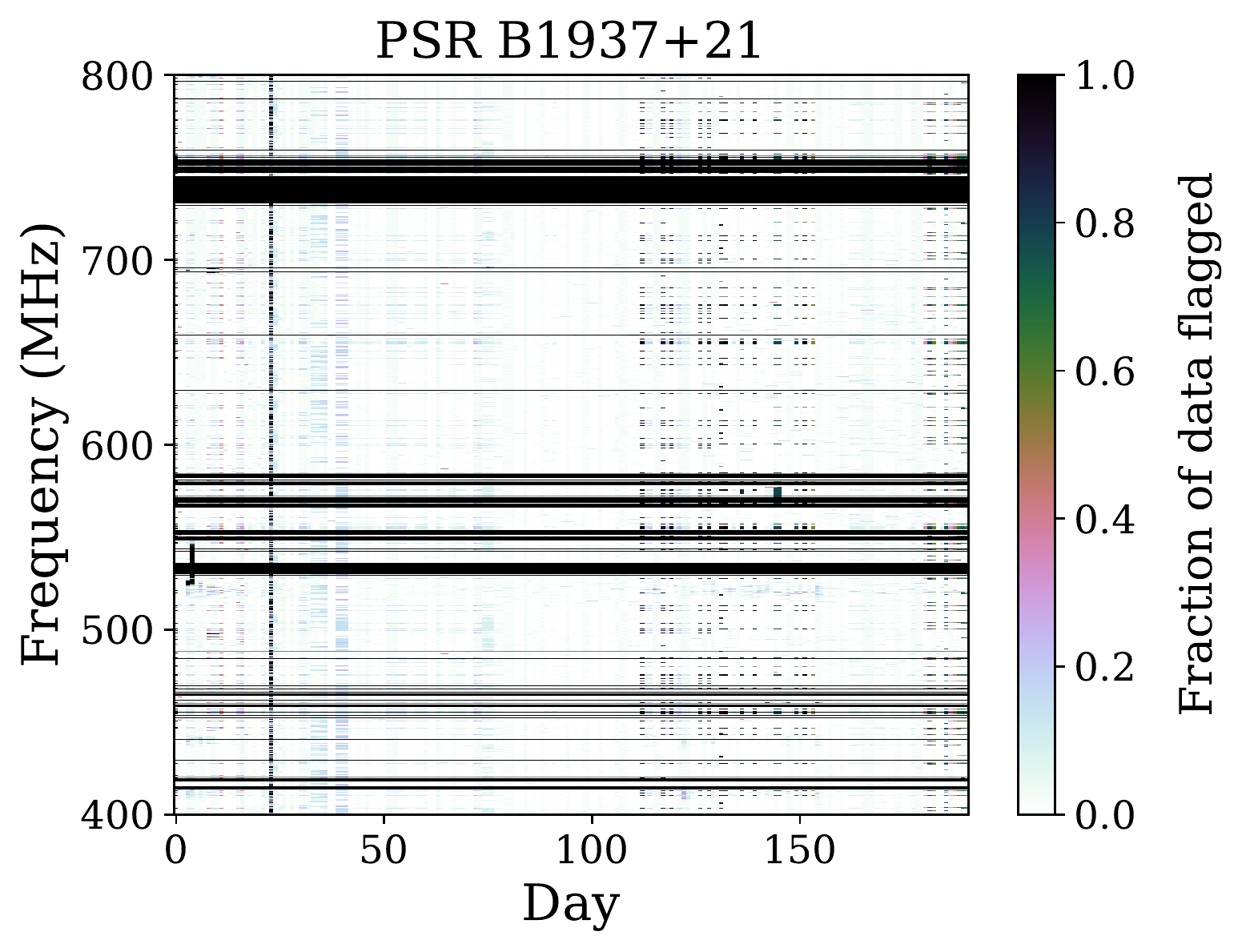}
\includegraphics[width=0.47\textwidth]{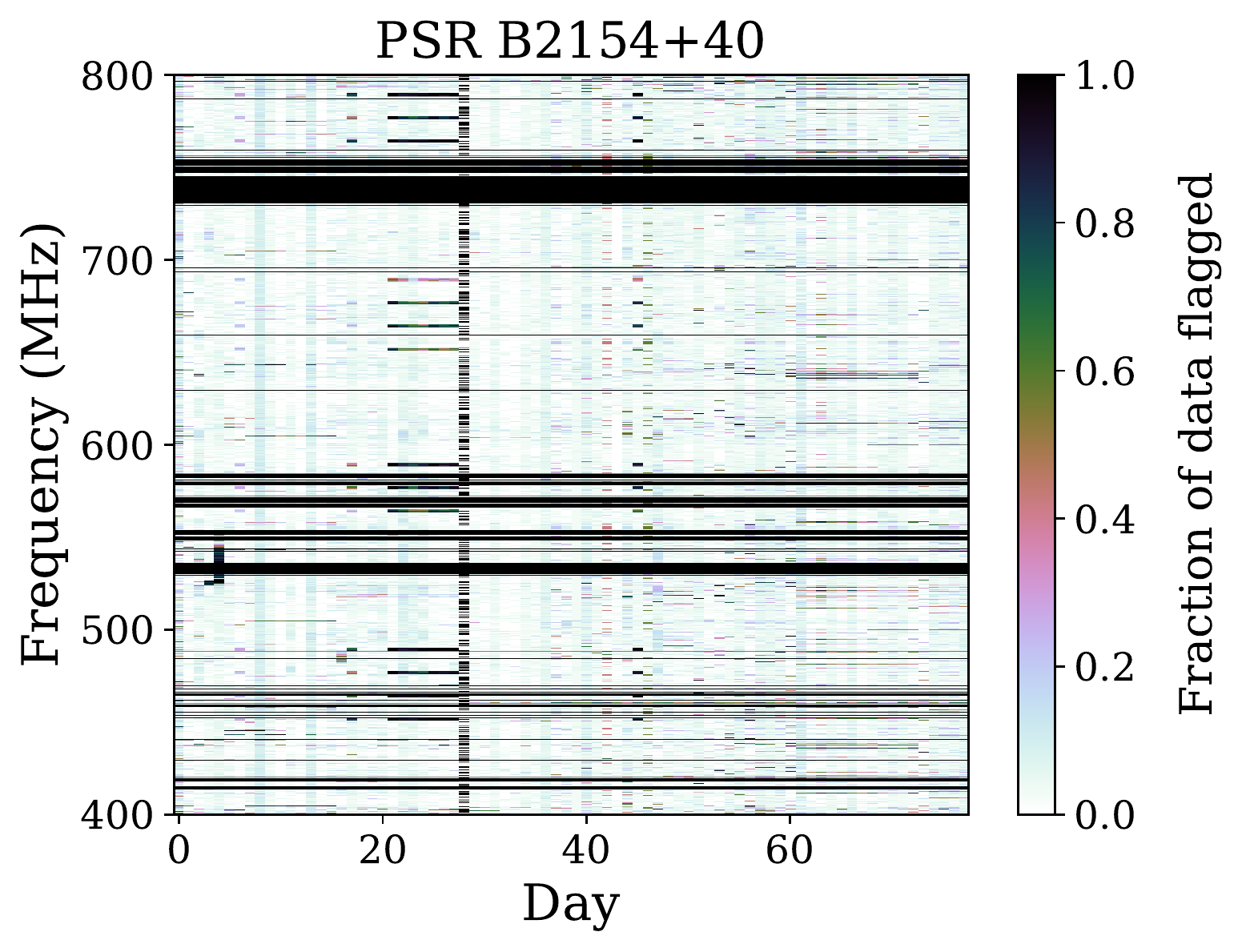}
\includegraphics[width=0.47\textwidth]{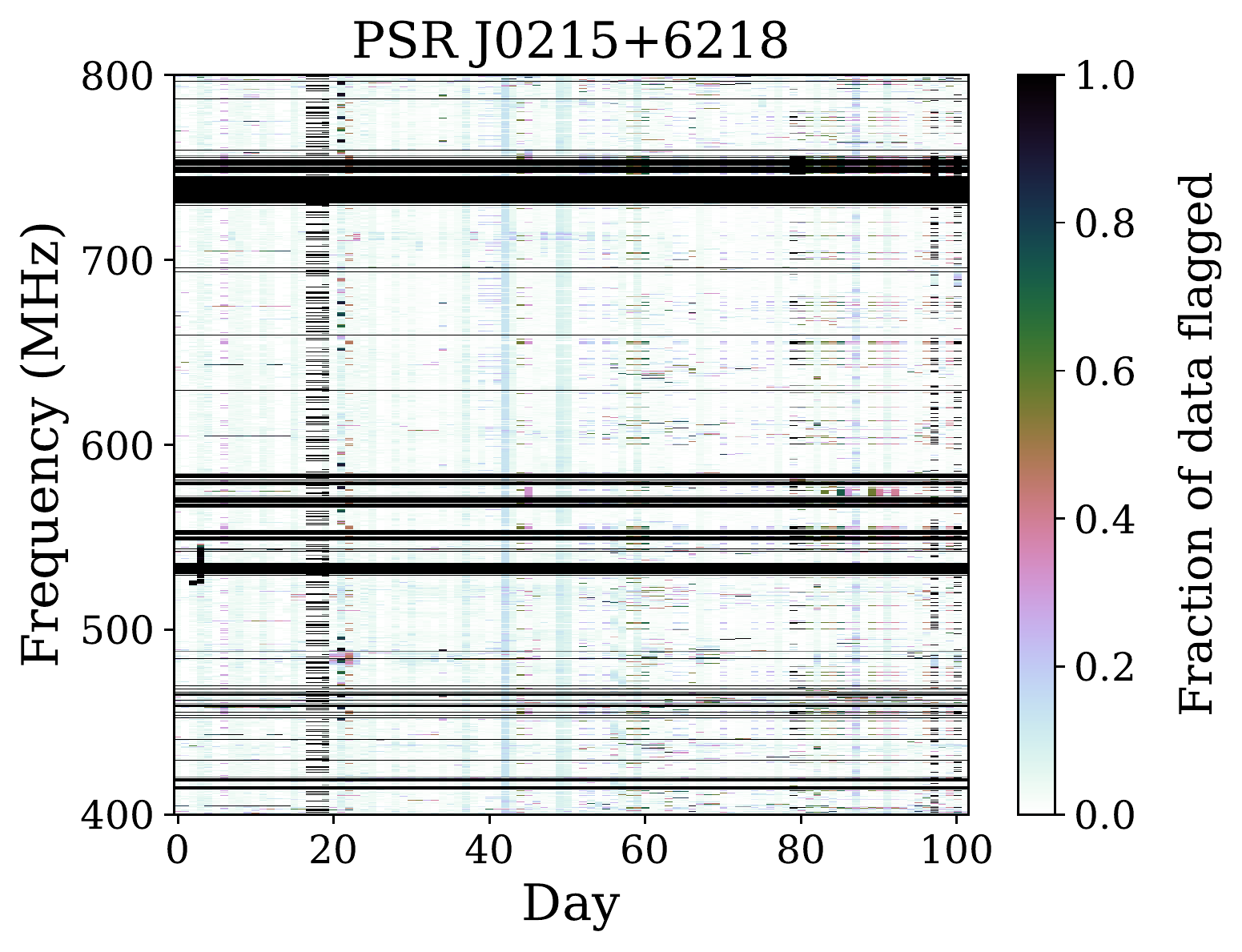}
\includegraphics[width=0.47\textwidth]{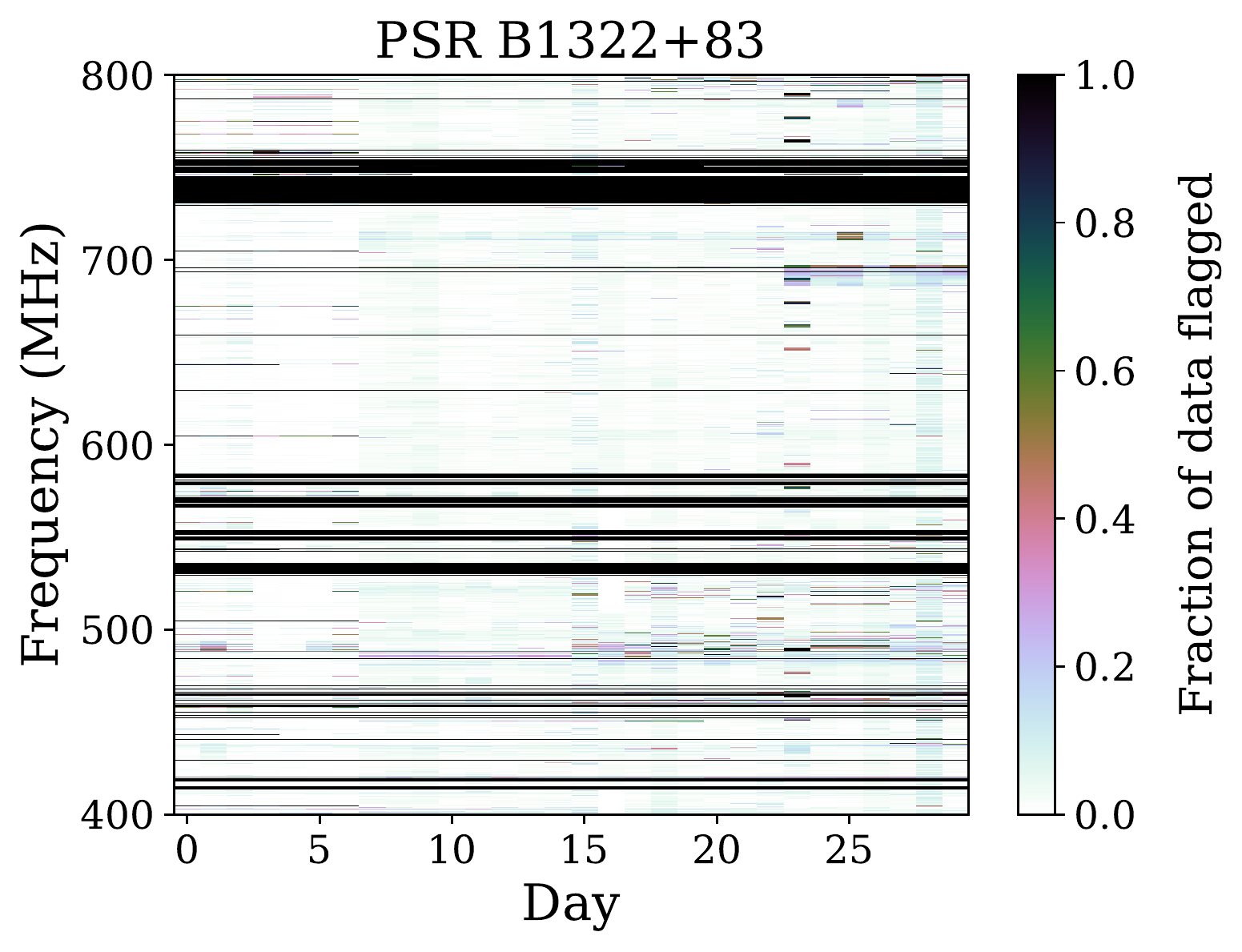}
\caption{A sample of post-processing RFI masks over time for six pulsars covering most of the \chimepsr{} observable declination range. Each observation was processed identically, which included applying a mask to known bad channels and an automated process based on RFI excision routines from the {\tt CoastGuard} software suite. The color bar represents the fraction of data flagged as RFI for each channel, for each observation. While individual observations (days) are presented in chronological order, the separation and total time spanned varies. The total time spanned for each RFI mask ranges from 164 to 324 days, and the range of observed dates is MJD 58500-58824 (2019-01-17 to 2019-12-07 UTC). The mean RFI fraction for all observations presented here is $\sim$20\% and the 95th percentile is $\sim$25\%, although from day-to-day the masking fraction can vary between $\sim$15\% to $\sim$68\%. \label{fig:multi_psr_badfreq}}
\end{figure*}

The bulk of RFI excision occurs during post-processing (e.g., while analyzing the folded or filterbank data products). A large fraction of the corrupted data can be removed by masking an empirically constructed list of frequency channels where RFI is persistent (approximately 15\% of the band). This corrupted channel mask can in principle also be combined with the calibration-based mask applied in the beamforming stage. More refined RFI excision can be attained by using techniques available in standard pulsar processing software packages, e.g., \psrchive{} or \presto{} \citep{presto}, but is left to user discretion. For example, in Figure~\ref{fig:multi_psr_badfreq} we have excised RFI using both the persistent corrupted channel mask and a method based on RFI cleaning utilities from the {\tt CoastGuard} software suite \citep{lkg+16}, where several robust statistical quantities on a per channel, per subintegration basis are evaluated to determine whether samples are corrupted\footnote{The exact version of the modified software used here can be found at \url{https://github.com/bwmeyers/iterative_cleaner/tree/v0.9}}. Typically, we mask $\sim$25\% of the 1024 frequency channels, yielding $\sim$300\,MHz of usable bandwidth spread non-contiguously across the full 400\,MHz observing band. As our understanding of the system and the RFI environment evolves, these will be tuned to improve sensitivity.

\subsection{Sensitivity}
CHIME is nominally capable of observing all known pulsars down to a declination of $\sim-20\degr$. 
To date, we have re-detected over $\sim$500 known pulsars with pre-commissioning \chimepsr{} observations. Out of these $\sim$500 observable sources, 209 pulsars have published flux densities at 600\,MHz in the \textit{ATNF Pulsar Catalogue}\footnote{\url{http://www.atnf.csiro.au/people/pulsar/psrcat}} \citep{mht+05}. 
For these pulsars, we can compare their detected average S/N with the expected value to assess the sensitivity of the \chimepsr{} system. 

We calculate the expected S/N for each pulsar using the radiometer equation. Many pulsars have steep spectral indices, and the sky temperature ($T_{\mathrm{sky}}$) will vary significantly across the 400\,MHz bandwidth. Therefore, in the estimation of the expected S/N, we separately consider three subbands and sum the three S/N values in quadrature. We take the flux density values at 400, 600, and 800\,MHz ($S_{400}$, $S_{600}$, and $S_{800}$ respectively) from the \textit{ATNF Pulsar Catalogue}. Where flux densities are not available, we extrapolate the subband flux density from the published spectral index. For each of the 209 sight-lines, we obtain $T_{\mathrm{sky}}$ at 400, 600, and 800\,MHz from the Haslam survey \citep{Haslam1982}. For the system temperature ($T_{\mathrm{sys}}$), we consider both the optimistic nominal value of 50\,K and a pessimistic worst-case-scenario of 100\,K to determine a range of expected S/N values.

\begin{figure*}
\centering
\includegraphics[width=0.8\textwidth]{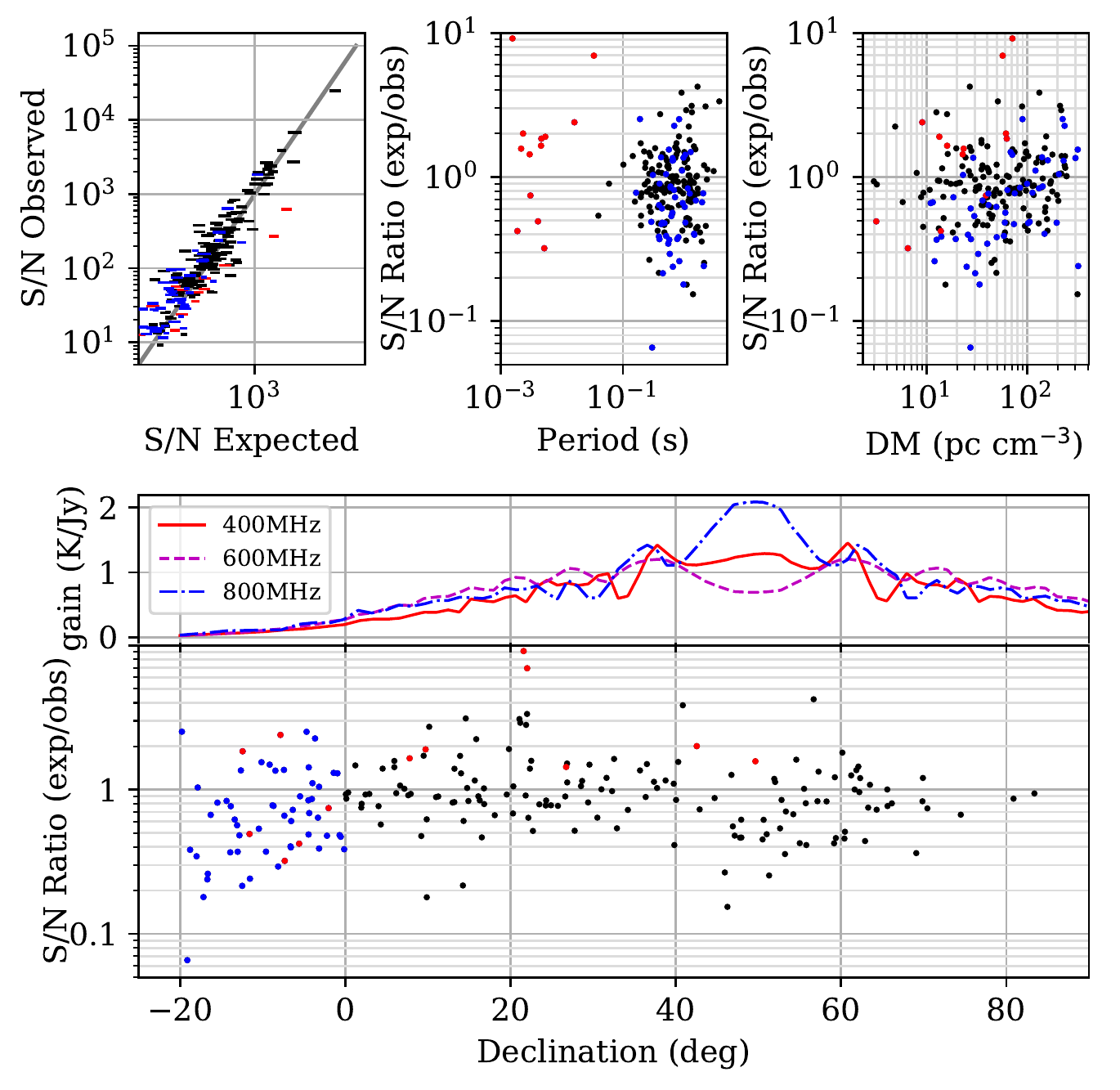}
\caption{A study of the expected versus observed S/N for 209 pulsars with published flux densities at 600 MHz. For each pulsar, the line-of-sight $T_{\mathrm{sky}}$ is estimated from the Haslam sky survey. In each panel, pulsars with declination $<0^\circ$ are highlighted in blue whereas those with short spin periods of $P<35\,$ms are highlighted in red. In the top-left panel we show the expected S/N is shown as a range for each pulsar, where the left edge of each point corresponds to a pessimistic $T_{\mathrm{sys}}$ of 100\,K and the right edge with a $T_{\mathrm{sys}}$ of 50\,K. The diagonal line is the 1:1 ratio, where the observed S/N equals the expected value. The ratio of the expected to observed S/N versus pulsar period (top-middle), and DM (top-right) are also shown. For simplicity, the expected S/N for each pulsar is taken to be the averaged value within the range of $T_{\mathrm{sys}}$. In the lower panel, we show the ratio of the expected to observed S/N versus declination, including the nominal CHIME primary beam response correction. The beam-corrected gain as a function of declination, at 400, 600 and 800\,MHz, is also provided.}
\label{fig:sensitivty}
\end{figure*}

Figure~\ref{fig:sensitivty} summarizes the results of this analysis at 600\,MHz. From the top left panel, it can be seen that most of our re-detections are close to the 1:1 diagonal line. The lowest published flux density of a pulsar we have re-detected is 0.6\,mJy. The CHIME primary beam response is strongly dependent on frequency and zenith angle, where at large zenith angles (low declinations, blue points in Figure~\ref{fig:sensitivty}) there is a significant decrease in sensitivity. Using a preliminary model, we have taken the primary beam attenuation into account when computing the expected S/N for pulsars, equalizing the response as a function of declination.

\subsection{Polarisation stability\label{sec:poln_stability}}
Precision pulsar timing requires exquisite polarisation calibration. In particular, the stability of the observed polarisation properties of any given pulsar over different epochs is critical \citep[e.g.,][]{vanstraten06,fkp+15}. A general polarisation calibration scheme for \chimepsr{} is under investigation and is a high-priority item for all CHIME projects.  Currently, however, we cannot transform from instrumental to intrinsic Stokes parameters. Nevertheless, Stokes profile stability can nominally be ensured by using arbitrary Mueller (transfer) matrices to effectively absorb any telescope miscalibration and depolarisation effects, including those introduced by averaging over frequency and time, when creating TOAs \citep[e.g.,][]{agh+18}\footnote{Original code is available here: \url{https://github.com/aarchiba/triplesystem/} in the \texttt{template\_matching.py} script.}. This approach is powerful and flexible, but it is not easy to translate the elements of the resulting transfer matrix to physical telescope properties.

\begin{figure}
    \centering
    \includegraphics[width=\columnwidth]{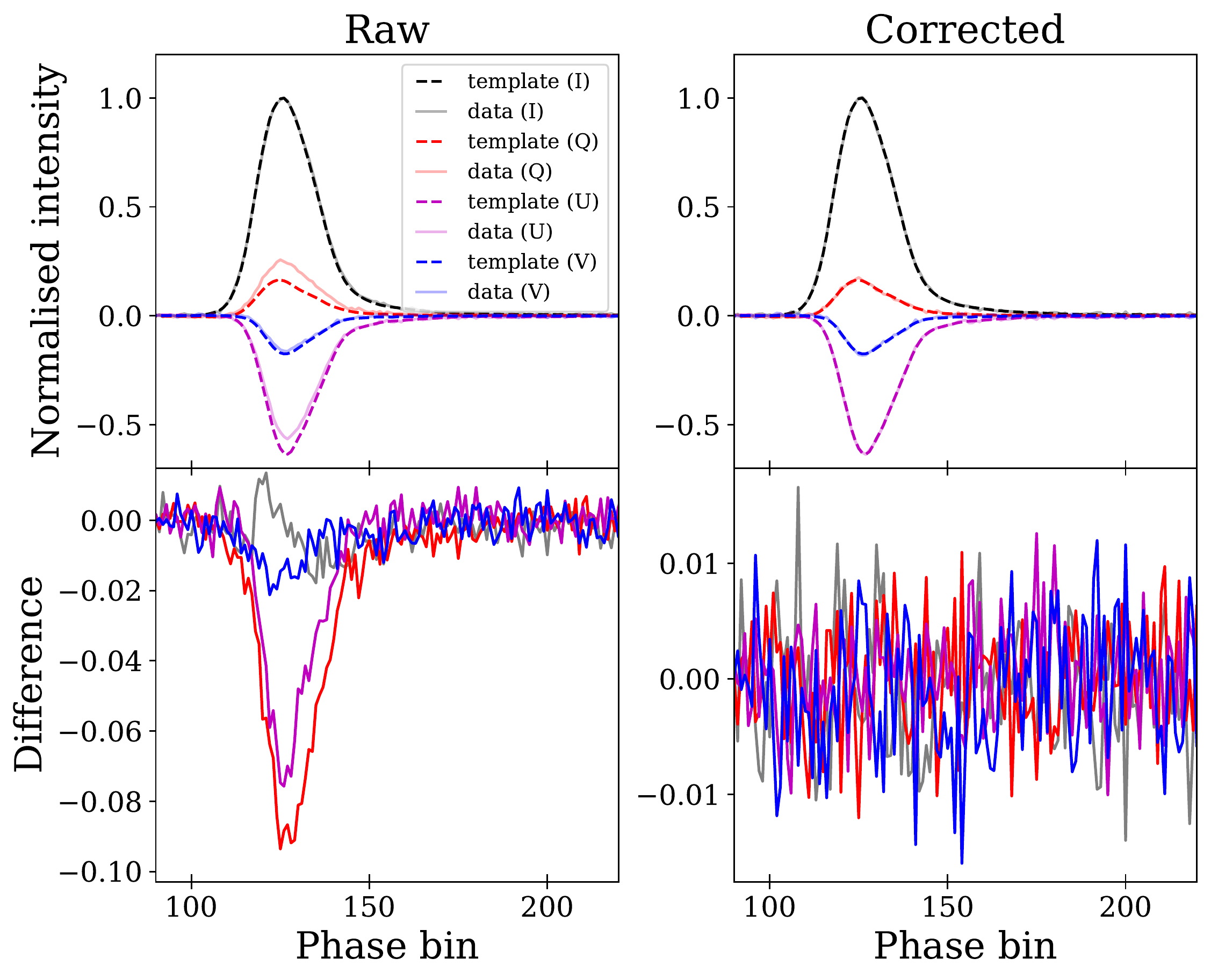}
    \caption{Comparison of the Stokes profiles from an observation of the main pulse of B1937+21 (solid lines) and a template (dashed lines). The top panels shows the uncorrected data (left) and corrected data (right) for each Stokes parameter: $I$ (black), $Q$ (red), $U$ (magenta) and $V$ (blue). The lower panels shows the residuals between the template and the raw data (left) and the corrected data (right).}
    \label{fig:stokes_stability_B1937}
\end{figure}

To examine the stability of the Stokes parameters over time, we inspected polarisation profiles of several bright targets, including B1937+21 (see Figure~\ref{fig:stokes_stability_B1937}). 
A high-S/N smoothed, frequency and time averaged template for each pulsar was created and used as a reference to compare observations of the same pulsar at different epochs. 
We compared the Stokes profiles and the resulting transfer matrix required to correct each observation to the reference profile, where the matrix elements should not differ substantially over time if the polarisation response of the system is stable.
The absolute difference between the template and observed Stokes profiles (without any correction), relative to the Stokes $I$ maximum, on different days is typically $\sim$10\%. 
Applying the arbitrary inverse transfer matrix corrections results in a residual difference of $\sim$1\% or less.
For individual pulsars, especially those which scintillate or exhibit mode changing, there are instances where the Stokes parameters change significantly on daily timescales relative to a given template.

Ultimately, polarisation stability and calibration is an on-going challenge for \chimepsr{}.
Applying the proposed arbitrary transfer matrix approach to stabilise the Stokes parameters from day to day will be a necessary step when conducting high-precision timing data analysis. 
Further effort will also be made into adapting this approach to a wideband regime, where the frequency dependence of the polarisation response is also considered.

\subsection{Spectral Leakage}
The polyphase filterbank (PFB) used by the CHIME FX correlator applies a sinc-Hann windowing function in the Fourier domain to raw digitized sky signal when forming 1024 frequency channels across the CHIME band \citep{baa+14}. While chosen to maximize channel sensitivity, the imperfect response in each synthesized channel produces leakage of signal between adjacent channels. Coherent dedispersion by the \chimepsr{} backend will ultimately produce aliasing of the detected pulsar signal, with a frequency-dependent lag between the original pulse and its aliased counterpart across the CHIME band.

\begin{figure}
    \centering
    \includegraphics[width=\columnwidth]{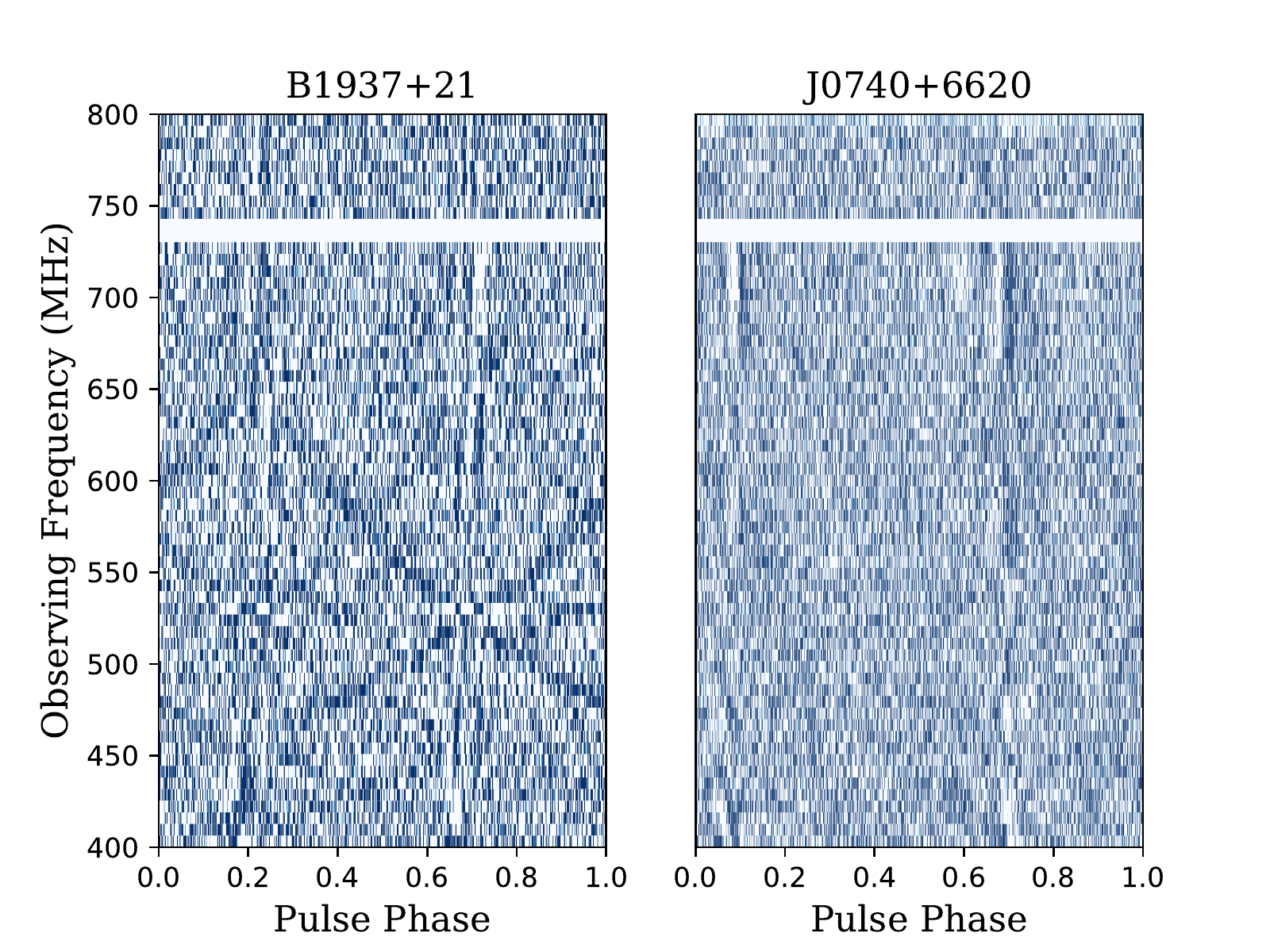}
    \caption{The presence of spectral leakage as dispersed features in \chimepsr{} spectra for PSRs J0740+6620 and B1937+21. In both panels, pulse profiles were removed using a principal component analysis for determining de-noised representations of the on-pulse dynamic spectrum. The spectrum for B1937+21 was obtained after integrating a single ($\sim$10-min) fold-mode recording, whereas the spectrum for J0740+6620 was determined by coherently averaging over 100 individual epochs of $\sim$20-min recordings. Vertical artifacts arise due to imperfect estimation of the de-noised template profiles.}
    \label{fig:spectral_leakage}
\end{figure}

Figure \ref{fig:spectral_leakage} shows examples of spectral leakage in pulse-averaged spectra for PSRs B1937+21 and J0740+6620. Artifacts due to leakage are apparent within a single observation of PSR B1937+21 with \chimepsr, though similar features are far less prominent in slower, fainter pulsars like J0740+6620, even after integrating data taken over 100 epochs as shown in Figure \ref{fig:spectral_leakage}.

The MeerTime pulsar-timing backend mitigates spectral leakage in the MeerKAT observing system by using a modified sinc-Hann windowing function in their F-engine system; the modified windowing function used by MeerTime suppresses the response of the synthesized-channel boundaries, which minimizes leakage but simultaneously lowers effective sensitivity in each channel \citep{bbb+18}. The commensal nature of pulsar/FRB observations with CHIME requires that the F-engine use the same PFB configuration for all backends when generating channelized data streams. Therefore, spectral leakage will be present to varying degrees in \chimepsr{} observations and its impact on timing will be assessed during offline processing.

\subsection{Timing}
In order to establish the timing capabilities of the \chimepsr{} system, we collected near-daily observations of MSPs observed by the North American Nanohertz Observatory for Gravitational Waves (NANOGrav) during a several-month commissioning period when telescope sensitivity was generally stable. We used timing solutions for online folding that are also used by NANOGrav at the 305-m Arecibo Observatory and the 100-m Green Bank Telescope (GBT) for pulsar-timing data acquisition. Initial ``template'' pulse profiles of \chimepsr{} data, necessary for cross-correlation and arrival-time estimation, were generated by excising channels containing RFI, coherently adding timing data taken across the commissioning period, and fully averaging the stacked set in time and frequency. The time- and frequency-averaged profile was then de-noised using a wavelet transform. 

Times of arrival (TOAs) were then computed via cross-correlation between the template profiles and a downsampled form of the \chimepsr{} fold-mode timing data. For analysis of NANOGrav pulsars, we fully integrated available fold-mode data in time and downsampled in frequency from the native resolution to 32 channels prior to TOA generation, yielding a maximum of 32 channelized TOAs per epoch. This level of downsampling is similar to the reduction methods used by NANOGrav in order to evaluate DM variations over time \citep[e.g.][]{lmj+16}.

\begin{figure}
\centering
\includegraphics[width=\columnwidth]{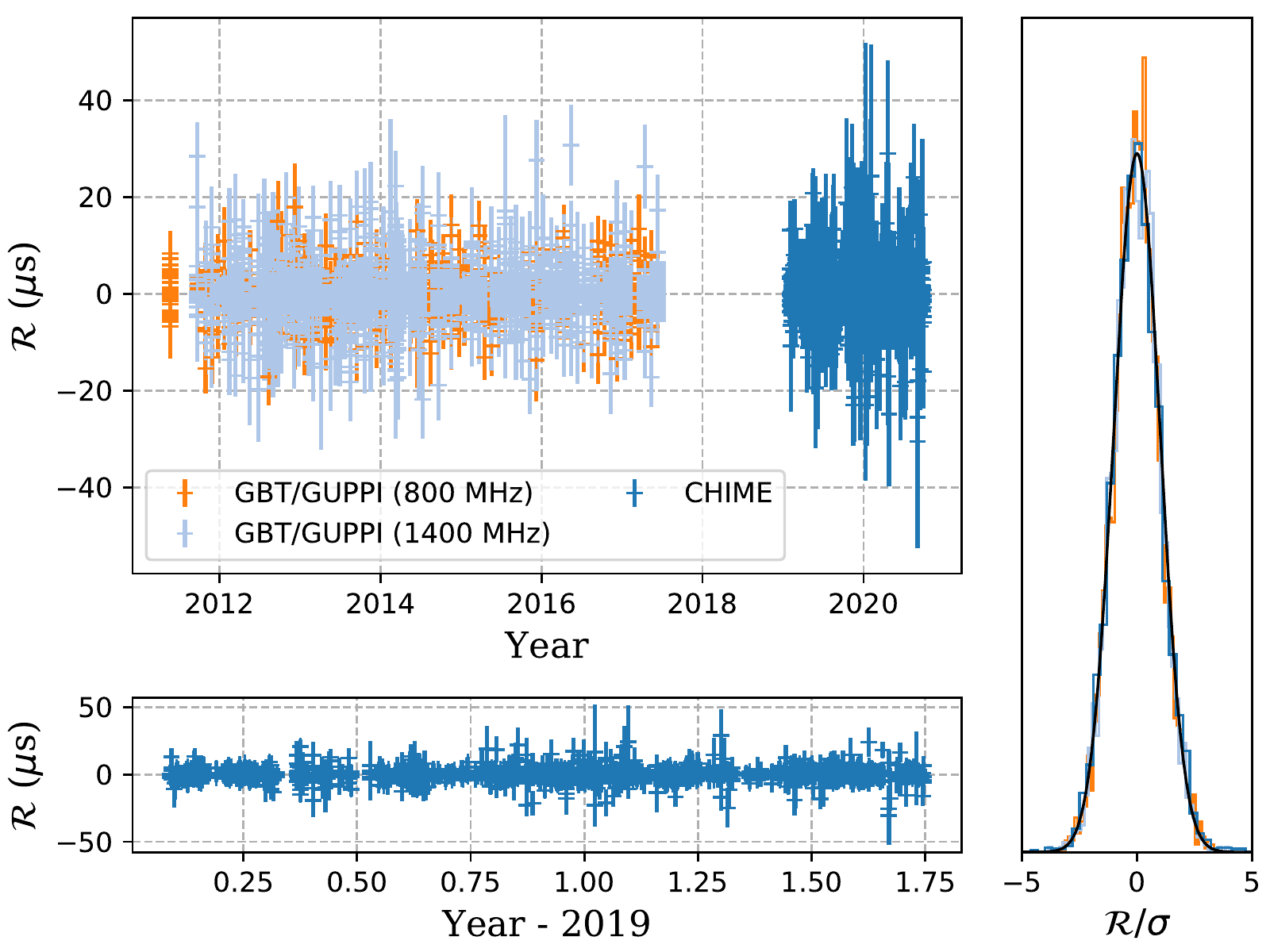}
\caption{A summary of best-fit timing residuals ($\mathcal{R}$) for PSR~J0645+5158. {\it Top-left.} The orange and light-blue points denote TOAs and best-fit estimates of $\mathcal{R}$ collected with the GBT using the 800-MHz and 1400-MHz receivers, respectively, as part of the NANOGrav 12.5-yr data release. Dark-blue points are TOAs and best-fit $\mathcal{R}$ measured from \chimepsr{} data. {\it Bottom-left.} A zoomed-in view of the \chimepsr{} timing data. {\it Right.} Normalized distributions of $\mathcal{R}$ weighted by $\sigma$, shown for all three TOA subsets as stepped histograms. The black line shows a unit-normal distribution for comparison, indicating that the \chimepsr{} system yields timing data with expected statistical properties and are consistent with TOAs obtained using other observatories.}
\label{fig:timing}
\end{figure}

Best-fit timing residuals for PSR~J0645+5158, computed using the \tempo{} pulsar-timing analysis package\footnote{\url{https://sourceforge.net/projects/tempo}} \citep{tempo} when modeling NANOGrav and \chimepsr{} TOA data simultaneously, are shown in Figure~\ref{fig:timing}. We used the timing solution generated for the 12.5-yr NANOGrav data release \citep{aab+20a,aab+20b} to model this combined data set, fitting for physical and instrumental parameters such as: pulsar-spin frequency and its first time derivative; astrometry (ecliptic coordinates, proper motion, and timing parallax); DM values estimated within one-day bins; and an arbitrary offset between the NANOGrav and CHIME data that reflects instrumental differences. As shown in Figure \ref{fig:timing}, we found that a numerically adequate timing model can be achieved when combining the CHIME and NANOGrav data sets, yielding best-fit parameters and fit statistics consistent with those previously reported from NANOGrav. The RMS residual for the CHIME data set on J0645+5158 is 0.83\,$\upmu$s, which compares well with the RMS values for the GBT data sets collected using the 800-MHz receiver (0.71\,$\upmu$s) and and 1400-MHz receiver (1.38\,$\upmu$s). A detailed timing study of NANOGrav sources observed with CHIME will be presented in a forthcoming study.

\section{Scientific Motivation}
\label{sec:science}

As described above, \chimepsr{} is a flexible system with a variety of acquisition modes that is designed for continuous and autonomous operation. 
Here we briefly outline some of the major scientific cases that will be explored by the \chimepsr{} project.

\subsection{Long-term Timing}
The long-term timing of pulsars, with a focus on PTA targets, is the primary science driver for \chimepsr{}. 
Detection of nanohertz-frequency gravitational waves using a PTA directly depends on a regular, high-cadence observing program of a large MSP sample over many years. 
The \chimepsr{} system is naturally producing such a data set with near-daily cadence. 
As such, \chimepsr{} timing data of NANOGrav sources \citep[e.g.][]{abb+18} will be combined in future NANOGrav data releases in order to improve detection of the stochastic GW background at nanohertz frequencies.

The autonomous design of the \chimepsr{} system, combined with the wide field of view of the telescope, allows for regular, priority-based timing observations of sufficiently bright pulsars, with 400--500 pulsars significantly detected within a single day of operation. This capability is unprecedented in the northern hemisphere and only recently achieved for southern pulsars
with the development of the MeerKAT pulsar timing program \citep[MeerTime;][]{bbb+18} and the UTMOST pulsar instrument \citep{bjf+17,jbv+19}. The \chimepsr{} system actively monitors observable sources that are reported in the ATNF pulsar catalogue with a range of cadence. \chimepsr{} also observes newly discovered pulsars found by the Green Bank North Celestial Cap \citep[GBNCC;][]{slr+14} survey in order to aid in the confirmation and follow-up of their discovered sources.

\subsection{High-cadence Timing of Pulsar-Binary Systems}
Long-term timing of binary radio pulsars often yields secular and/or periodic variations from purely Keplerian motion. The most famous examples of secular variations are those associated with general-relativistic orbital decay and precession in compact orbits \citep[see][for a review]{sta03}, though many non-relativistic pulsar orbits have been observed to vary over time due to evolving sky orientations induced by proper motion \citep[e.g.][]{kop95,kop96}. By contrast, the relativistic Shapiro time delay \citep{sha64} is a periodic effect observed in sufficiently inclined binary systems of any size where the pulsed signal traverses varying amount of spacetime curvature induced the companion star over the course of the orbit. Measurements of such effects allow for direct constraints on the masses \citep[e.g., PSR~J0740+6622;][]{cfr+20} and geometry of the systems in question \citep[e.g., PSR~J0437$-$4715;][]{vbb+01}, and thus stand to yield high-impact information that is otherwise inaccessible from purely Keplerian dynamics.

\begin{figure}
\centering
\includegraphics[width=\columnwidth]{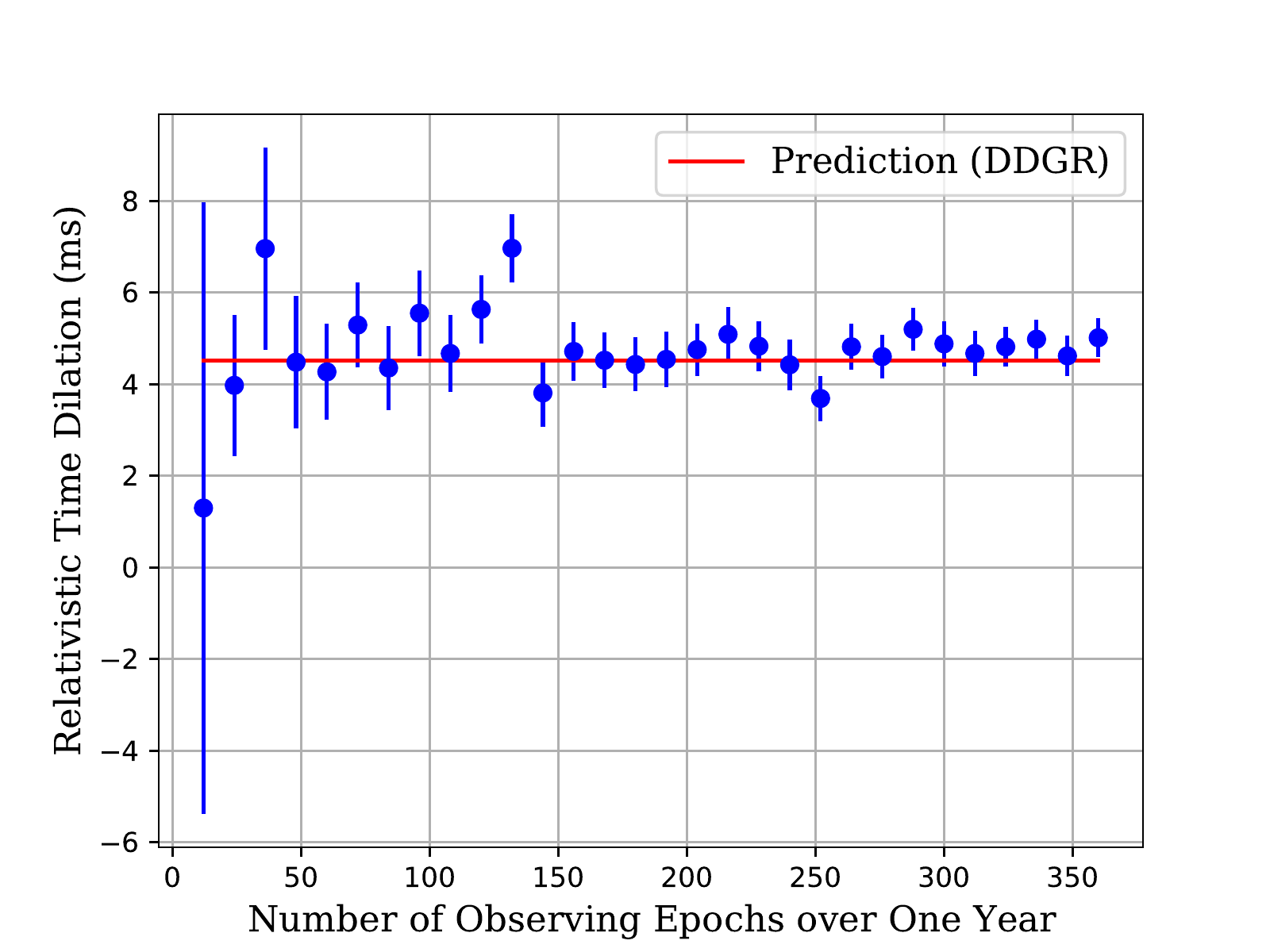}
\caption{Estimates of the relativistic parameter quantifying time dilation and gravitational redshift for PSR J0509+3801 -- typically referred to as $\gamma$ in pulsar-timing literature -- derived from simulated TOA data sets with TOAs collected over one year in time but with different observing cadences. For each plotted measurement, a timing data set is generated with {\tt tempo} assuming that one frequency/time-averaged TOA is obtained per epoch, and that the TOA data set yields white-noise properties consistent with current \chimepsr{} observations of J0509+3801 (i.e., epoch-averaged RMS residual of $\sim$70 $\upmu$s). The red horizontal line marks the expected value determined by \citet{lsk+18} when modeling their TOAs using the ``DDGR" timing model, that assumes all variations are effects predicted by general relativity.}
\label{fig:time_dilation}
\end{figure}

The \chimepsr{} system is expected to produce a rich data set for probing binary astrophysics by monitoring all visible binary pulsars with near-daily observing cadences over the course of telescope operation. As an example, Figure~\ref{fig:time_dilation} shows simulated best-fit estimates of a timing parameter quantifying relativistic time dilation and gravitational redshift in the PSR J0509+3801 double-neutron-star system \citep{lsk+18}. While typical programs observe such sources on monthly or bi-monthly cadences, the simulated estimates in Figure~\ref{fig:time_dilation} suggest that such effects can be better constrained with high-cadence observations -- like those achievable with \chimepsr{} -- due to quickened evaluations of orbital parameters and their variations. Moreover, we anticipate the daily cadence to yield detections of the Shapiro delay on timescales much shorter than those typically seen in current pulsar-timing literature, due to the faster rate of achieving dense orbital coverage.

\subsection{Plasma propagation effects}
\label{sec:plasma_propagation_effects}
Pulsars are sensitive probes of the ISM and its structural variations across many lines of sight. By processing and recording data at low radio frequencies, the \chimepsr{} system is producing a rich and growing data set for autonomously monitoring frequency-dependent features in pulsar data. Examples of such effects include temporal variation of DM \citep[e.g.][]{lcc+16,lkd+17,leg+18}, frequency-dependent DM \citep{css16,dvt+19}, ``echoes" in pulsar spectra \citep[e.g.][]{glj11,mhd+18,djb+19,bts+20}, ``extreme scattering events" in flux density modulations \citep[e.g.][]{cks+15,kcw+18}, scintillation \citep[e.g.][]{brg99a,brg99b,brg99c,wmj+05,wym+08} or multi-path scattering/pulse broadening \citep[e.g.][]{bcc+04,mls+18} caused by small-scale plasma structures. Understanding these effects is critical in achieving a GW background detection using PTAs \citep[e.g.][]{cs10,lmj+16}, but are also informative when considering pulsar astrometry and system dynamics \citep[e.g.][]{lyne+84,pmd+14,rch+19}.

\begin{figure}
\centering
\includegraphics[width=\columnwidth]{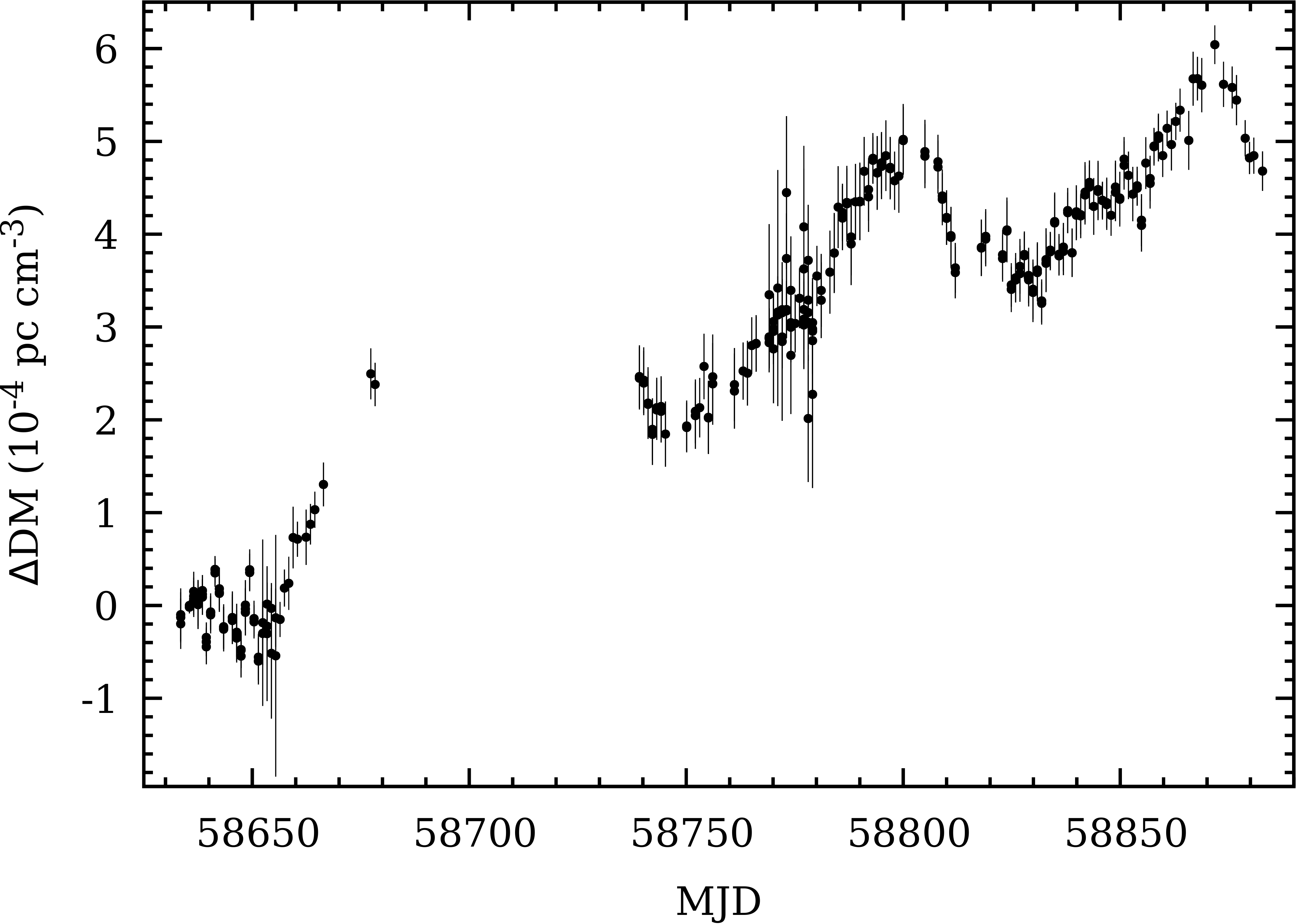}
\caption{DM time series measured from CHIME observations of PSR\,B1937+21, relative to the initial value of $71.01572$\,\dmunits{}. Each observation was integrated in frequency to 16 channels, with a TOA measured per channel. The DM was determined by fitting for the frequency-dependent dispersive delay between in-band TOAs for each day. The  median DM uncertainty is $2.9\times10^{-5}$\,\dmunits{}. The mean cadence following MJD\,58742 is 1.2\,days.\label{fig:B1937DM}}
\end{figure}

As a demonstration of the suitability of \chimepsr{} for studying small-scale dispersive variations, we measured the DM time series for PSR~B1937+21, one of the fastest-known MSPs, and one that is known to exhibit large and rapid variations in DM (e.g \citealp{jml+17}), following the process detailed in \cite{dvt+19}. The 10 highest-S/N observations were selected, which were phase-aligned and summed to create a high-S/N frequency-resolved reference profile, with very little correlated noise. The original 1024 frequency channels were integrated to 16 channels, and a wavelet smoothing algorithm was applied, resulting in a noise-free standard template. The remaining observations in our data set were also integrated to 16 frequency channels, and TOAs were measured for each via cross-correlation with the template. The TOAs were analysed using \tempotwo\footnote{\url{https://bitbucket.org/psrsoft/tempo2/src/master}} \citep{hem06}, with a timing model based on the one presented in \cite{pdd+19}, with the time-varying DM and noise models removed. We fit for DM on each observing epoch while keeping the other timing-model parameters fixed to obtain a time series with a mean cadence of 1.2\,days and a median DM precision of $2.9\times10^{-5}$\,\dmunits{}, which we present in Figure~\ref{fig:B1937DM}. As a comparison, the DM time series for B1937+21 obtained from NANOGrav observations with the Arecibo and Green Bank observatories yield a similar mean DM precision of $\sim$2$\times10^{-5}$\,\dmunits{}, though evaluated over a monthly cadence \cite{aab+20a,aab+20b}. The measured DM time series from \chimepsr{} data displays an overall linear trend, with additional short-duration excesses lasting a few tens of days, which would not be easily-resolved in low-cadence data sets (such as those normally employed by PTA experiments).

Another important PTA science consideration that will benefit from the \chimepsr{} observing campaign is the study of the Solar wind, which contributes an annual variation in DM to pulsar timing data. 
This effect varies in time, throughout the course of the 11-yr Solar cycle, and it has been proposed (e.g. \citealp{mca+19}, \citealp{tvs+19}) that long-term, high-cadence radio monitoring of pulsars close to the ecliptic plane will provide valuable information about this phenomenon. 
Recent work has shown that in high-cadence data sets with high DM precision, typical simple Solar wind models do not adequately-describe the excess dispersive delay \citep{tvs+19}, and that its robust mitigation will become increasingly important in searches for low-frequency gravitational waves as pulsar timing array sensitivity continues to improve \citep{mca+19}.

\subsection{Polarization}
Polarization properties of radio pulsars have provided a wealth of information about their emission mechanism and geometry, and about the magnetised Galactic medium \citep[e.g.][]{lk04}. 
The \chimepsr{} backend records and stores full Stokes information, allowing polarization studies to be performed for all detected sources. 
However, polarization calibration, necessary for many pulsar studies, is difficult to obtain for a transit telescope such as CHIME, where the interplay between the two orthogonal sets of feeds changes during the observation and is strongly frequency dependent. 
Obtaining a beam model of the telescope accurate enough to recover the intrinsic polarization of the signal is a work in progress involving measurements from various CHIME backends, including \chimepsr{} (see also Section~\ref{sec:poln_stability}). 
Nevertheless, even before being able to calibrate the instrument, some interesting properties of polarized sources can be measured, and in particular the effect of Faraday rotation. 
\chimepsr{}, operating at low frequencies and with a large fractional bandwidth, is able to provide very precise measurements of rotation measure (RM).

The capability of \chimepsr{} to measure precise RM values has already been demonstrated in a recent work by \citet{ng20_rm}, where we present RM values for 80 pulsars, 55 of which are measured for the first time. 
We plan to measure the RM of most detectable northern pulsars, many of which either do not have a reported RM value, or the catalogued value was measured decades ago. 
By virtue of observing a large fraction of northern pulsars regularly, \chimepsr{} will also significantly contribute to efforts to map the Galactic magnetic field, complementing similar ongoing studies at low frequencies \citep[e.g.][]{lml+18,sbg+19}. 
In addition, frequent observations present an exciting opportunity to investigate temporal RM variations of every observed source. 
Measuring simultaneous DM variations, as described in \textsection\ref{sec:plasma_propagation_effects}, will enable the measurement of Galactic magnetic field gradient.

\subsection{Glitch monitoring}
Approximately 200 pulsars to date\footnote{See \url{http://www.jb.man.ac.uk/pulsar/glitches/gTable.html} and \url{http://www.atnf.csiro.au/research/pulsar/psrcat/glitchTbl.html} for a list of glitches and their corresponding publications.} have been observed to exhibit ``glitches", where their rotation period abruptly changes and can remain altered for several weeks or months subsequently \citep[e.g.][]{rm69,wje11,mjs16,lbs20}. 
Studying glitches and their aftermath provides insight into the interior structure of neutron stars and the nuclear equation of state \citep[e.g.][]{hm15,hs18}. 
The modular design the \chimepsr{} system lends itself to immediate, offline analysis of timing variations consistent with pulsar glitches from many different sources, and potentially enables near-realtime alerting of such events to the transient-astronomical community. 
Such analyses are especially interesting in the era of LIGO/VIRGO detection of GWs, where correlations of glitch and GW transient events are expected to yield multi-messenger constraints on the interior structure of neutron stars.
Furthermore, the observing cadence of the pulsar with a detected glitch can be increased in order to better model the post-glitch recovery properties of the pulsar.

\subsection{Monitoring for emission variability in pulsating sources}
Daily observations, flexible recording configurations, and a wide observing band are ideal for long-term monitoring of the emission from objects such as repeating FRBs \citep[e.g.][]{ssh+16,chimefrb_second_repeater,chimefrb_repeaters19,fab+20}, RRATs \citep{mll+06}, intermittent \citep{ksm+06}, nulling \citep{b70}, and mode-changing pulsars \citep{b70b,l71,njm+17}. 
This will allow us to probe any short and long-term variation in the rotational properties of the sources and their potential link with the variability in emission properties \citep[e.g.][]{lhk+10,psw+16,slk+19,njm+18}.

The high-cadence observations also allow us to search for previously unknown emission variability in known pulsars. 
As a testament to this, \citet{ng20_nulling} identified new nulling and mode changing pulsars from the commissioning data set (July 2018 to March 2019).

\subsection{Searching for pulsars, RRATs, pulses from repeating FRBs and transient astronomical events}
The filterbank mode of the \chimepsr{} backend is being actively used to follow up both unknown single pulses of Galactic origin (Good et al., submitted) and repeating FRB sources discovered by CHIME/FRB.
The large instantaneous field of view and the transit nature of CHIME/FRB has allowed for the blind detection of these sources.
A number of these sources are then followed up with the more sensitive \chimepsr{} backend as the data are coherently dedispersed to the DM of the sources, allowing detection of faint pulses that will otherwise be missed by CHIME/FRB. 
Moreover, the tracking beams generated for \chimepsr{} remain centered on the positions of the sources during their transit at the CHIME site, allowing for robust detection of the intrinsic dynamic spectrum.
Furthermore, \chimepsr{} has a time resolution of 327.68\,$\upmu$s, three times higher than CHIME/FRB, that could resolve smaller scale structures of the single pulses otherwise undetectable by CHIME/FRB.

Follow-up tracking observations of the unknown Galactic sources with \chimepsr{} is also used to look for periodicity akin to pulsars and RRATs.
These are done by either measuring the gaps between successive pulses, or periodicity searches including both Fast Fourier Transform and ``fast folding algorithms"~\citep[e.g.][]{pkr+18}.
The new candidates detected by the CHIME/FRB system and confirmed by \chimepsr{} are subsequently monitored for long-term timing\footnote{See \url{https://www.chime-frb.ca/galactic}}. 
New Galactic discoveries from the use of both CHIME/FRB and \chimepsr{} is reported in \cite{gac+20}.

While individual observations of a source with \chimepsr{} are limited by the primary beam size of CHIME which restricts transit time, the capability to perform multiple simultaneous observations means that we can afford to repeatedly observe a number of sources daily to stack and search the data for pulsations. 
This will allow for the potential detection of faint sources by leveraging long integration times. Such strategy has been employed before in targeted searches of globular clusters \citep[e.g.][]{crf+18}.

\chimepsr{} is also able to rapidly follow up on any transient astronomical events that might emit radio pulses. 
The regular schedule can be overwritten quickly to point a pulsar beam towards the source of interest during the forthcoming transit. 
We expect to apply the capability to follow up on events such as magnetars undergoing X-ray outburst \citep[e.g.][]{crh+06,rep+13,erb+20}, gravitational wave events \citep{aaa+16,aaa+17} and new transitional MSPs \citep{asr+09}.

As with fold-mode observations, filterbank scans of known or candidate sources are assigned priorities and included in the automatic scheduling of acquisition with \chimepsr{}. 
However, it is possible to instead program \chimepsr{} to operate as a dedicated blind-search machine. 
In this case, the FX correlator and \chimepsr{} backend can beamform and process, respectively, up to $\sim$3000 distinct lines of sight per day for 5-minute acquisition times.

\section{Conclusions \& Future Directions}\label{sec:conclusions}
In this work, we reported on a system built for the CHIME telescope to enable radio pulsar observations in different modes of acquisition. 
We described the hardware setup of the \chimepsr{} backend and provided details on the software/networking modules that were developed to interact with the CHIME FX-correlator for generating 10 synthesized tracking beams based on autonomous source selection and scheduling. 
The \chimepsr{} system operates in tandem with the cosmology and FRB experiments, and is capable of regularly recording data in both ``fold" and filterbank modes of pulsar observations. 
The \chimepsr{} system is poised to make a wide variety of important science contributions ranging from detailed studies of known sources and discoveries of new ones, to investigations of the ISM, both on its own and in collaboration with other instruments.

\acknowledgements{
We are grateful to the staff of the Dominion Radio Astrophysical Observatory, which is operated by the National Research Council of Canada.  CHIME is funded by a grant from the Canada Foundation for Innovation (CFI) 2012 Leading Edge Fund (Project 31170) and by contributions from the provinces of British Columbia, Québec and Ontario. The CHIME/FRB Project, which enabled development in common with the CHIME/Pulsar instrument, is funded by a grant from the CFI 2015 Innovation Fund (Project 33213) and by contributions from the provinces of British Columbia and Québec, and by the Dunlap Institute for Astronomy and Astrophysics at the University of Toronto. Additional support was provided by the Canadian Institute for Advanced Research (CIFAR), McGill University and the McGill Space Institute thanks to the Trottier Family Foundation, and the University of British Columbia. The CHIME/Pulsar instrument hardware was funded by NSERC RTI-1 grant EQPEQ 458893-2014. We thank Erik C. Madsen for his work in planning the development of the CHIME/Pulsar backend. We also thank the anonymous referee for their feedback and suggestions, which improved the quality of this work. Pulsar research at UBC is funded by a NSERC Discovery Grant and by CIFAR. This research was enabled in part by support provided by WestGrid (\url{www.westgrid.ca}) and Compute Canada (\url{www.computecanada.ca}). V.M.K. holds the Lorne Trottier Chair in Astrophysics \& Cosmology and a Distinguished James McGill Professorship and receives support from an NSERC Discovery Grant and Herzberg Award, from an R. Howard Webster Foundation Fellowship from the CIFAR, and from the FRQNT Centre de Recherche en Astrophysique du Quebec. D.M. is a Banting Fellow. S.M.R. is a CIFAR Fellow and is supported by the National Science Foundation (NSF) Physics Frontiers Center award 1430284. The National Radio Astronomy Observatory is a facility of the NSF operated under cooperative agreement by Associated Universities, Inc. P.S. is a Dunlap Fellow and an NSERC Postdoctoral Fellow. The Dunlap Institute is funded through an endowment established by the David Dunlap family and the University of Toronto. C.L. was supported by the U.S. Department of Defense (DoD) through the National Defense Science \& Engineering Graduate Fellowship (NDSEG) Program. M.D. receives support from a Killam fellowship, NSERC Discovery Grant, CIFAR, and from the FRQNT Centre de Recherche en Astrophysique du Quebec.}
\\
\software{
\dspsr{} \citep{DSPSR}, \psrdada{} (\url{http://psrdada.sourceforge.net}), \psrchive{} \citep{PSRCHIVE_hvm04, PSRCHIVE_vdo12}, \presto{} \citep[][]{presto}, \tempo{} \citep{tempo}, \tempotwo{} \citep{hem06}
}

\appendix 

\section{Observed pulsar list}
We collected observing statistics for all pulsars from the ATNF Pulsar Catalogue over a representative time period, MJD 59100--59274, where the scheduling algorithm described in Section \ref{subsec:scheduler} was executed with no modifications. This information is provided in the accompanying machine-readable table, and a small example is given in Table~\ref{tab:obs_list}. The table columns are described as follows:
\begin{itemize}
    \item[]\textit{Column 1} (\texttt{psr\_name}): The pulsar name.
    \item[]\textit{Column 2} (\texttt{start\_mjd}): The MJD corresponding to the first observation of the pulsar with \chimepsr{}.
    \item[]\textit{Column 3} (\texttt{avg\_obs\_per\_week}): The mean number of observations per week for the pulsar.
    \item[]\textit{Column 4} (\texttt{avg\_snr\_per\_obs}): The mean time- and frequency-averaged S/N for a typical observations of the pulsar.
\end{itemize}

\begin{table}
\centering
\caption{The first 15 rows from the machine-readable table of observed sources. \label{tab:obs_list}}
\begin{tabular}{cccc}
\hline
(1) & (2) & (3) & (4)\\
PSR & Start MJD & Number of obs. & Mean S/N\tablenotemark{a,b} \\
 & & per week\tablenotemark{a} &  \\
\hline \hline
J0002+6216 & 59033 & 2.8 & 4 \\
J0006+1834 & 58441 & 6.4 & 12 \\
J0011+08 & 59031 & 5.9 & 11 \\
B0011+47 & 58170 & 6.3 & 129 \\
J0023+0923 & 58440 & 6.6 & 23 \\
J0026+6320 & 58441 & 6.3 & 67 \\
J0030+0451 & 58443 & 5.8 & 37 \\
J0033+57 & 58413 & 6.6 & 4 \\
J0033+61 & 59179 & 6.3 & 13 \\
J0034$-$0534 & 58439 & 5.3 & 15 \\
B0031$-$07 & 58170 & 4.7 & 118 \\
J0034+69 & 59030 & 7.0 & 4 \\
J0039+35 & 59032 & 3.7 & 5 \\
B0037+56 & 58170 & 5.6 & 189 \\
B0045+33 & 58170 & 2.8 & 78 \\
\hline
\end{tabular}
\tablenotetext{a}{Mean values were calculated over a representative 3-month period.}
\tablenotetext{b}{S/N values for each observation were computed after integrating over transit time and observing frequency.}
\end{table}

\bibliographystyle{aasjournal}
\renewcommand{\bibname}{References}

\bibliography{references}

\end{document}